\documentclass[aps,amsfonts, twocolumn,prd,nofootinbib]{revtex4}
\usepackage{graphicx}
\usepackage{url} 
\usepackage[hyperindex]{hyperref} 
\hypersetup{
    bookmarks=true,            
    unicode=false,          
    pdftoolbar=true,        
    pdfmenubar=true,       
    pdffitwindow=true,     
    pdftitle={A Class of Nonperturbative Configurations in Abelian-Higgs Models: 
Complexity from Dynamical Symmetry Breaking},    
    pdfauthor={Marcelo Gleiser and Joel Thorarinson},    
    pdfsubject={Extended Field Configurations},       pdfnewwindow=true,     
    pdfkeywords={EFC, PRD, Theory, Nonequilibrium Field Theory},
    colorlinks=true,         
    linkcolor=blue,                  
    citecolor=blue,     
    filecolor=magenta,      
    urlcolor=blue,
    breaklinks=true,
    hyperfigures=true,          
}
\newcommand{\bra}{\langle}
\newcommand{\ket}{\rangle}

\newcommand{\im}{\mbox{Im}\,}
\newcommand{\re}{\mbox{Re}\,}
\newcommand{\Tr}{\mbox{Tr}\,}

\def\raw { \rightarrow  }
\def\lb { \left\{    }
\def\rb { \right\} }

\def\La       {{Langevin~}}

\def\L       {{\mathcal L}}
\def\D       {{\mathcal D}}

\def\pd      { \partial}
\def\V       {{\mathcal V}}

\def\z       {{\zeta}}
\def\L       {{\mathcal L}}
\def\H       {{\mathcal H}}

\def\pd { \partial}
\def\V       {{\mathcal V}}
\def\O       {{\mathcal O}}
\def\bra     {{\langle}}
\def\ket     {{\rangle}}

\def \dx {\delta x}
\def \dt {\delta t}

\def \a {\alpha}

\def\b{\beta}

\def\mn { {\mu\nu}}

\def\be{\begin{equation}}
\def\ee{\end{equation}}
\newcommand{\eea}{\end{eqnarray}}
\newcommand{\bea}{\begin{eqnarray}}

\def\g{\gamma}
\def\d{\delta}
\def\l{\lambda}
\def\d{\delta}

\def\l{\lambda}

\def \p { \phi}
\def \r { \rho}
\def \P { \Phi}

\def\pdbk{\bra \phi^\dag \ket}
\def\pbk{\bra \phi \ket}

\def\V2{\mu^2 \pdbk \pbk + \lambda (\pdbk \pbk )^2 }

\def\A       {{\mathcal A}}
\def\D       {{\mathcal D}}

\def\hx       {{\hat{x}}}

\def\xmm {{\hat{x}-\hat{\mu}}}
\def\dag {{\dagger}}

\def\mdag {{\mu \dagger}}

\begin{document}
\title{A Class of Nonperturbative Configurations in Abelian-Higgs Models: 
Complexity from Dynamical Symmetry Breaking}
\author{M.~Gleiser}
\email{gleiser@dartmouth.edu}
\author{J.~Thorarinson}
\email{joel.thorarinson@gmail.com}

\affiliation{Department of Physics and Astronomy, Dartmouth College,
Hanover, NH  03755, USA\\ \\}

\begin{abstract} We present a numerical investigation of the dynamics of symmetry breaking in both Abelian and non-Abelian $[S U (2)]$ Higgs models in three spatial dimensions. We find a class of time-dependent, long-lived nonperturbative field configurations within the range of parameters corresponding to type-1 superconductors, that is, with vector masses ($m_v$) larger than scalar masses ($m_s$). We argue that these emergent nontopological configurations are related to oscillons found previously in other contexts. For the Abelian-Higgs model, our lattice implementation allows us to map the range of parameter space -- the values of $\beta = (m_s /m_v )^2$ -- where such configurations exist and to follow them for times $t \sim \O(10^5) m^{-1}$. An investigation of their properties for $\hat z$-symmetric models reveals an enormously rich structure of resonances and mode-mode oscillations reminiscent of excited atomic states. For the $SU(2)$ case, we present preliminary results indicating the presence of similar oscillonic configurations. 
\end{abstract}
\maketitle


\section{Introduction}
The mechanism of spontaneous breaking of local gauge symmetries plays a fundamental role in our current understanding of high-energy particle physics \cite{Weinberg1996} and of condensed matter systems \cite{Zinn-Justin1989}. In very general terms, it can be stated that a spontaneously broken symmetry is always associated with the existence of degenerate vacuum states: the theory predicts the existence of discrete or continuously degenerate vacua while Nature chooses one of them. Strictly speaking, the mixed matrix elements describing possible transitions between vacuum states only vanish in the infinite-volume limit. However, since mixed matrix elements scale with volume ${\cal V}$ as $\exp[-c{\cal V}]$, where $c$ is a positive constant, for all practical purposes symmetries do get broken for large enough volumes. 

In the realm of relativistic quantum field theories, the particle spectrum of a given model is computed as small perturbations about a broken-symmetric vacuum state. In addition to these, models with nontrivial nonlinear couplings may also have nonperturbative, solitonic solutions to their equations of motion \cite{Rajamaran1987}. These usually come in two kinds: for models with nontrivial vacuum structure, there can exist static solutions that owe their stability to the topology of the vacuum manifold, the so-called topological defects \cite{Kibble1976,Vilenkin1994}. For example, models with a $1d$ (one spatial dimension) real scalar field with a double-well potential have kink solutions, while Abelian-Higgs models in $2d$  have Nielsen-Olesen vortices. In both cases, symmetry is restored at the core of the topological defect. In contrast, nontopological solitons owe their stability to the conservation of a global charge \cite{Friedberg1976,Coleman1985}. The distinctive signature of both topological and nontopological solitons is their time-independence: they are effectively static solutions to the equations of motion, even in the case of $Q$-balls, where the complex scalar field is written as $\P\sim \exp[-i\omega t]$ so as to transform the time-dependent term in the equation of motion to a mass term $\sim \omega^2\P^2$. 

Given the vast richness of spatiotemporal phenomena in Nature \cite{Cross1993}, one should suspect that other possible nonperturbative configurations exist in relativistic field theories once we allow for their time-dependence. Usually, these are not taken into account, as they are expected to be short-lived and hence dynamically uninteresting. However, if long-lived configurations do exist, they are bound to play a crucial role when fields are far from equilibrium and, in particular, during symmetry breaking. If present, they will change our understanding of the vacuum, as they comprise, together with possible topological and nontopological extended field configurations (EFCs), nonperturbative fluctuations about it: for example, if we are to sum over possible contributions to the path integral, these must be included. In a cosmological setting, if we are to study the approach to thermalization during post-inflationary reheating \cite{Micha2004,Bassett2006}, these configurations may possibly be very important \cite{Farhi2008,Graham2006}. 

As more recent research has shown, there are abundant examples of such long-lived solutions.\cite{Gleiser1994,Copeland1995,Farhi2005,Hindmarsh2006,Fodor2006a,Graham2007,Gleiser2007,Saffin2007}. The first hint was the discovery of breathers in $1d$ kink-antikink scattering (see, e.g., ref. \cite{Campbell1983}): for certain relative velocities a new time-dependent configuration, a breather, would form. Remarkably, breathers were never seen to decay. Their demise may only occur through highly-suppressed nonperturbative decay modes. Even before that, spherically-symmetric configurations called ``pulsons'' were found by Bogolyubsky and Makhankov \cite{Bogolyubsky1976,Makhankov1978}. These were the first examples of the configurations later called oscillons in ref. \cite{Gleiser1994,Copeland1995}, where it became clear that such oscilating large-amplitude real scalar field solutions are present in any $3d$ model with amplitude-dependent nonlinearities as long as the potential has a region of negative concavity, $V'' < 0$. For more details on the properties of scalar field oscillons see ref. \cite{Gleiser2008}.

The more recent extension of oscillons to the Standard Model presented in refs. \cite{Farhi2005} and \cite{Graham2007} prompted us to search for oscillons in the Abelian-Higgs model. In our first work \cite{Gleiser2007}, we found that vortex-antivortex annihilation in $2d$ can, for a range of parameters, generate remarkably long-lived oscillon-like configurations. These are characterized by a see-saw oscillation in the $z$-component of the magnetic field and seem to owe their stability to a gauge-field induced mass gap for the scalar field, although an in-depth study is still lacking.
 
In the present work we explore the existence of oscillon-like states in $3d$ Abelian-Higgs models. We find that they not only exist but are quite easily found in the context of type-1 superconductors, that is, for gauge-field masses substantially larger than Higgs-field masses. As in $2d$, we construct an effective phase diagram mapping the range of parameters where oscillons are found. Using extrapolation, we obtain an approximate critical value of the control parameter $\beta=(m_s/m_v)^2$ beyond which we conjecture that no oscillons are produced during symmetry breaking. 

The crucial point, though, is that these configurations {\it emerge spontaneously} during symmetry breaking. We don't start with an approximate spherically-symmetric solution and see it relax into oscillons, as has been the rule in such studies. The $U(1)$ oscillons literally condense dynamically, as the system transitions from its symmetric to its broken-symmetric state. Although it is true that we take advantage of the formation of flux-tubes to facilitate the formation of oscillons -- a $3d$ analog of the vortex-antivortex annihilation in $2d$ -- those flux tubes occur naturally during the symmetry-breaking process. It is quite remarkable that, in a model where there is no topologically-stable defect, flux-antiflux tubes annihilation will form long-lived EFCs. Even though they do not show the same symmetry restoration at the core as do topological defects, we show that in $3d$ $U(1)$ oscillons clearly probe into the $V''<0$ part of the potential. We also briefly discuss the remarkably rich structure of these objects, which display resonant mode-mode fluctuations reminiscent of excited atomic states. Finally, we present preliminary results in the context of $SU(2)$ Higgs models where, although we were not able to find long-lived configurations, there is a strong indication that they exist in the type-1 regime.

This paper is organized as follows: in section II we present the model and our conventions. In section III we describe our search for $U(1)$ oscillons and how we managed to isolate them to investigate their longevity. In section IV we present the phase diagram for $3d$ $U(1)$ oscillons and obtain the critical value of the control parameter $\b$. In section V we describe the rich resonant structure of these configurations, stressing the similarities with excited atomic states. In section VI we present our preliminary results for type-1 $SU(2)$ models. In section VII we present our conclusions and a brief summary of our results. Finally, the two appendices describe the technical details and issues of the lattice implementation of Abelian and non-Abelian models. Particular attention is paid to the proper implementation of gauge constraints in the presence of stochastic forcing terms.


\section{$U(1)$ \& $SU(2)$ Equations and Conventions} \label{section:equations}

We use the continuum Lagrangian,

\be 
\label{eq:Lag}
\L= \D_\mu \P^\dag \cdot \D^\mu \P-\frac{1}{4}F^{\mu \nu} \cdot F_{\mu
\nu}+ \frac{1}{4}\l(\P^\dag \cdot \P-\eta^2)^2, 
\ee 
where $\D_\mu = \pd_\mu +i g A_\mu$, and $A_{\mu}$ is either the Abelian $U(1)$ or non-Abelian $SU(2)$ gauge field. The context within which we will be using one or the other will be clear. The operation $\cdot$ does nothing for $U(1)$ but is equal to $a\cdot b \equiv \frac{1}{2} \Tr a^\dag b$ for $SU(2)$ matrices. Performing the scaling $A_\mu\rightarrow
\eta^{-1}A_\mu$, $\phi\rightarrow \eta^{-1}\phi$ and $x\rightarrow
\eta g x$, there is only one independent parameter in these models, the ratio of scalar to vector masses, $\beta \equiv (m_s/m_v)^2 = \l/(2g^2)$. For convenience, we will keep $\l=2$ and vary the gauge coupling $g$ so that all times are quoted in units of the scalar mass, $[t]=m_s^{-1}$.

Note that this choice of conventions can be related to condensed matter models that use the hopping parameter $\kappa$, (for example, see reference \cite{Alford:2007np}) by $\kappa\equiv 1/\sqrt{2} g$. The critical parameters separating type-$1$ (small $\b$) and type-$2$ (large $\b$) for $U(1)$ superconductivity is $\b=1$ and $\kappa=1/\sqrt{2}$.

The continuum equations for this system are,
\be
 \D_\mu \D^\mu \P =
\pd_{\P^\dag} V ; 
\ee  
\be \D_\mu F^{\mu \nu}=  J^\nu \label{fmunu_} ~,
\ee  
where $J^{\mu} = i\left (\P^{\dagger}D^{\mu}\P - \P D^{\mu}\P^{\dagger}\right )$ is the conserved current. We will solve these equations numerically in the temporal gauge $A_0=0$, which makes the time component of the last equation a nondynamical constraint. This gauge also allows the definition of simple functions for the conjugate momenta and hence for a Hamiltonian lattice implementation. Details of the lattice implementation can be found in the Appendices. 


\section{Finding Oscillons in the $U(1)$ Theory}  \label{section:findingosc}

In a previous work, we showed that in $2d$ it is possible to find long-lived, time-dependent oscillon configurations from the annihilation of vortex-antivortex pairs (henceforth vav) \cite{Gleiser2007}. These remarkable EFCs are characterized by a persistent see-saw oscillation of the magnetic field $B_z$ and very little emitted radiation. [The interested reader can see an animation {\tt \href{http://media.dartmouth.edu/~mgleiser/U1Q5FCircle.avi}{here}}.] We were also able to show that the formation of $U(1)$ oscillons can be described as a phase transition in field configuration space with $\b$ as the control parameter: oscillons were shown to form from vav annihilation only for $\b\leq \b_c\simeq 0.13(6)\pm 2$. We constructed an order parameter $E_{\rm osc}/E_v$, where $E_{\rm osc}$ is the oscillon energy and $E_v\simeq 2\pi\b^{1/5}$ is the Nielsen-Olesen vortex energy, showing that
$E_{\rm osc}/E_v\sim \left (|\b - \b_c|\right )^{0.2(2)\pm 2}$.

\begin{figure} 
\includegraphics[scale=.6,angle=0]{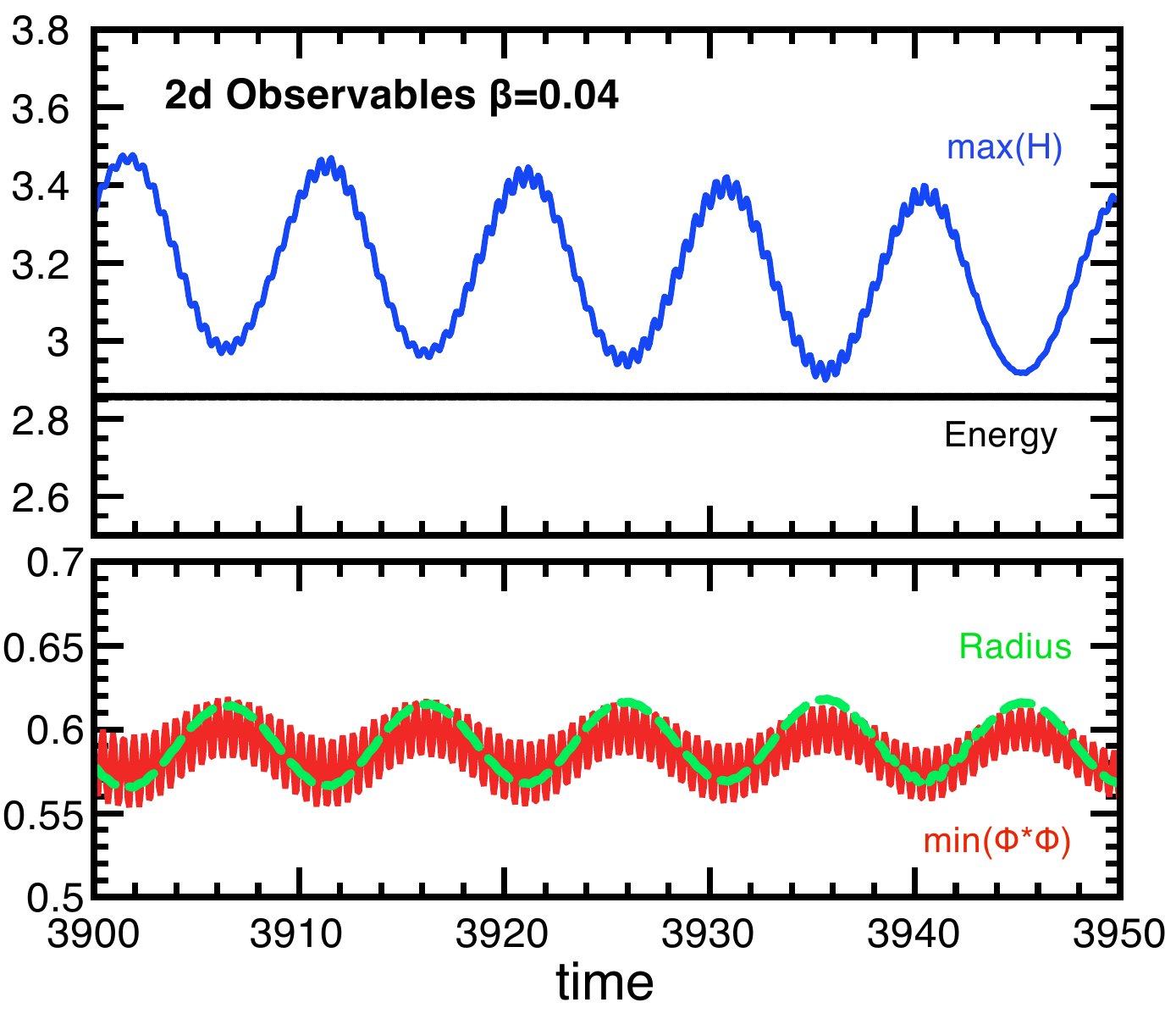}
 \caption{\label{fig:2d_oscil} (Color online) A few observables characterizing the $2d~ U(1)$ oscillon. The blue line is the maximum energy density of the EFC while the black line is the total energy within a radius of $R=4$. The lower plot shows both the expected radius of the EFC $R_{eff}$ (green) and the minimum value of $\P^*\P$ within the EFC (red). All of these observables except the energy are oscillatory in time and have a nontrivial fourier space distribution of frequencies.}
\end{figure}

Figure \ref{fig:2d_oscil} shows a few relevant observables during the oscillon phase. Note the constancy of the total energy (black line), integrated over a finite volume surrounding the oscillon. The blue line denotes the local maximum of the total Hamiltonian density, max$[{\cal H}(t,x,y)]$. The program scans the lattice at each time step to find the local maximum of the energy density. The dashed green curve displays the value of the effective radius of the configuration, computed as $R_{eff}^2\equiv\frac{\int_0^{r_0} r^2 \H(r) dr}{{\int_0^{r_0} \H(r) dr}}$, with $r_0\sim {\rm few~} m_s^{-1}$. The red curve denotes the minimum amplitude of the scalar field, ${\rm min}[\P^{\dagger}\P(t,x,y)]$. Here, it is important to notice that the vacuum is at $\P^{\dagger}\P |_{\rm vac}=1$, while the inflection point is at $\P^{\dagger}\P |_{\rm inf}=0.5$. Thus, the $2d$ oscillon is marginally nonperturbative as the field amplitude hovers just around the inflection point. We will see that this will not be the case in $3d$.

It is natural to search for similar oscillon-like EFCs in $3d$. However, as soon as this is attempted, one meets a few obvious challenges. First, contrary to $2d$, there are no stable topological vortices in $3d$. We can find $U(1)$ flux tubes, but they do not share the stability of $2d$ Nielsen-Olesen vortices. Second, it is computationally much harder to search for EFCs in $3d$. As we will see, if we start from a symmetric initial state and quench to the broken symmetric state, there will be an excess of energy that makes potential candidates harder to isolate.

Generally, finding an oscillon involves some method of confining enough energy within a small spatial region and then seeing if the energy remains localized without any or with very little dissipation to infinity. This method implies that oscillons can be thought of as attractors in field configuration space: through their natural dynamical evolution, the interactions among the fields will conspire to create an oscillon solution. This property of oscillons has recently been demonstrated analytically in the context of pure real scalar field models \cite{Gleiser2008}. A similar proof for gauge models is lacking, but our numerical results indicate that this property will also hold for these more realistic models.

From our experience with the $2d$ theory we know that for small $\beta=m^2_s/m^2_v$, oscillons can form from vav decay. We can try to extend this result to $3d$. Note that the fact that oscillons form easily from vav annihilation for small $\b$ doesn't mean that they {\it only} form for small $\b$: we have observed that they can also form from a zero-phase Gaussian initial condition in $\P$ for larger $\b$, $\P =1-\P_0\exp[-r^2/R^2]$, where $\P_0$ is the departure from the vacuum at $|\P|^2=1$ and $R$ its spatial extension. 

We stress that there is a fundamental difference in these two approaches. Using a Gaussian as an initial condition relies on our knowledge of the approximate oscillon behavior and is equivalent to coaxing the solution into existence by letting the fields relax into it from a nearby point in field configuration space. This is commonly done in numerical relaxation techniques. A quench, on the other hand, doesn't assume any initial profile for the solutions: the oscillons emerge spontaneously as the system works to minimize its energy and maximize its entropy dynamically. Thus, finding oscillons through this dynamical approach offers strong evidence of their existence as attractors in field configuration space. It means we should expect them to be present during symmetry breaking.

We now describe the procedure to form oscillons from vav annihilation. We first thermalize the fields in a symmetric single well potential at low temperature $T$ and with viscosity $\g =1$. We do this using the stochastic Langevin approach described in the Appendices. We then switch to the double well potential, while at the same time turning off the stochastic noise but maintaining the viscosity. The viscosity will dampen excessive energy and will allow a more transparent identification of the many vortices and anti-vortices formed on the lattice. In steps:

\begin{itemize}
\item Following the lattice implementation in appendices 1 and 2, set all fields ($\A_{\mu},\P$) and their derivatives to zero. 
\item Thermalize using \La dynamics in a quadratic potential with minimum at $\P= 0$. The temperature $T$ should be chosen to generate a distribution of fluctuations across $k$-space. A typical number is $T\sim 0.1m_s$. Using $\g=1$ (in units of $m_s$), the system should thermalize well within $t\sim 10 m^{-1}_s$. We take the system to be thermalized when $\pd_t\bra \Pi^2\ket \sim 0$, where $\Pi=\pd_t\P$ and the brackets denote a volume average. 
\item Switch to $T\raw 0$ while keeping $\g=1$ and simultaneously switch to the double well (Higgs) potential.
\item Then evolve for $t\sim 15m^{-1}$ until the vortices and antivortices form.
\item Set $\g=0$, and then evolve conservatively.
\end{itemize}

Nearby vortices (anti-vortices) will then interact to form higher-$N$ vortices (anti-vortices) while nearby vav pairs annihilate. In $2d$, for strongly type-$1$ parameters, ($\beta \leq 0.136$) the process of vav annihilation has a good probability of forming gauged oscillons \cite{Gleiser2007}. Whether they form this nontopological, time-dependent, radially-symmetric bound state depends on both the size of the perturbative modes around the structure and on the relative velocity of the vortices as they interact. Our strategy in $2d$ was to minimize as much as possible their relative velocity using the viscosity. We also noticed the robustness of the formed oscillons against perturbative radiative modes.

\subsection{Attempt 1 in $3d$: Gaussian {\it ansatz}}

In $3d$ the dynamics is significantly different, as the flux-tubes are not topologically stable as are vortices in $2d$. Also, there are more decay paths, and more surface energy to compensate for in higher dimensions. We will attempt both the Gaussian ansatz initial condition and the stochastic field quench as we did in $2d$ to search for oscillon-like structures. As an aside, we note that our approach to find new time-dependent nonperturbative solutions is applicable to a variety of nonlinear partial differential equations. The same way that oscillons could never have been predicted with the perturbative analytical techniques commonly used to study nonlinear equations, we conjecture that many nonperturbative time-dependent solutions remain unknown. Small-amplitude, spatially-extended oscillons were recently found in the context of real scalar field theories {\it a posteriori}, after they were numerically discovered. These objects are amenable to a small amplitude treatment, as has been shown in refs. \cite{Farhi2008} and \cite{Fodor2008}.

\begin{figure} 
\includegraphics[scale=.34,angle=0]{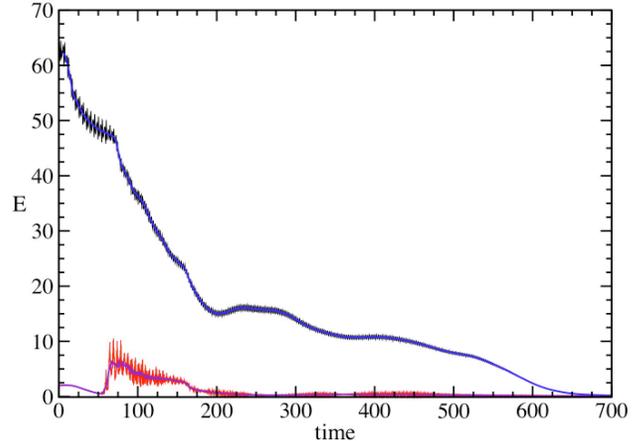}
 \caption{\label{fig:3dq5} (Color online) Attempt to find a $3d$ $U(1)$ oscillon from a Gaussian initial condition for $\b=0.04$ (or $g=5$). Plotted is total energy (blue, upper curve)  and magnetic field energy (red, lower curve) within a $10^3$ box. Note that, although the configuration is unstable, the leaking of energy is much slower than expected in a linear theory. }
\end{figure}

In $3d$, the first result we report is a negative one. In figure \ref{fig:3dq5} we plot the total energy (blue, upper curve) and its contribution from gauge fields (red, lower curve) for a Gaussian perturbation from the vacuum. The ansatz is $\P=1- \P_0 e^{\frac{r^2}{R^2}}$, with $\P_0=1$ and $R=4$ and for $\b=0.04$. No information was put in the gauge fields, that is, we took $A_i=0$. We tried many different combinations of radii, amplitudes and couplings $g$, but none of the configurations were stable. Even though this same type of ansatz was successful in the equivalent $2d$ theory for a large range of $\beta$, it is clearly too simplistic for $3d$. As we know that $3d$ $U(1)$ oscillons exist (see below), a longer-lived oscillon should be found with the ansatz method, although one would need to incorporate the gauge fields in a more sophisticated way. As we are more interested in the dynamical emergence of oscillons, we will not pursue this further. It is worth mentioning that although the configuration decays  in $t \leq 700m_s^{-1}$, this is still enough time for the object to dramatically affect macroscopic physics. It is important to keep this in mind when judging the possible implications of short lived nonperturbative resonances.

\subsection{Attempt $2$ in $3d$: Breaking the Symmetry}

Our next attempt involved making a string network and letting those strings interact to see if long-lived, localized nontopological spatiotemporally-complex objects were formed. To do this, we followed the same steps outlined above: first, we thermalized the field to $T_{\rm latt}=T\dx^{-d}$ in a quadratic potential; second, we quenched it by switching to the double-well potential to seed the formation of strings; third, we maintained the viscosity long enough so that the strings stabilized and the scalar field approached its vacuum expectation value. The viscosity dynamically decreased the temperature $T\raw ~0$. As in $2d$, the only consideration when choosing an initial $T$ is to generate enough excited modes to seed the formation of a sufficient density of strings upon symmetry breaking while not so high that they are overdense. Specific parameters are given below. Finally, after the string network formed, the friction was turned off and the strings interacted. We then had to search for nontrivial structures resulting from these interactions. We started to explore at very small $\b$, since in $2d$ that is when structures are more likely to form. Also, we expect $\b_c$ to be smaller than in $2d$, as there will be more surface tension to compensate for. In what follows, we describe the details of two searches, characterized by the initial parameters as configurations 1 and 2, respectively. This should allow our results to be reproducible by other groups.

\subsubsection*{Configuration $C_1$}

The details of configuration $C_1$ were: 
\bea 
\lb d, N_L,L,\dx,\dt, g,\l,\eta,T_{\rm latt}  \rb_{C_1} \equiv \\ \lb 3,96,19.2,0.2,0.05,4,2,1,0.25 \rb, \nonumber 
\eea 
where the various symbols stand for: $d$ -- spatial dimensionality; $N_L$ -- number of lattice points; $L$-- lattice length; $\delta x$ -- lattice spacing; $\delta t$ -- time discretization; $g$ -- gauge coupling; $\l$ -- scalar coupling; $\eta$ -- viscosity; $T_{\rm latt}$ -- lattice temperature. 
As remarked, the basic procedure for constructing this configuration is similar to how we formed vortices in $2d$, although we have more sensitive dependence on the time-scales for string formation in $3d$. For clarity, the steps are:
\begin{itemize} \label{recipe_c_1}
\item Set all fields to their vacuum configurations, $\P\sim 0$ in the quadratic potential $V=\P^\dagger \P$.
\item Thermalize using \La dynamics with $\g=1$ for $t\sim 10 m_s^{-1}$ at $T_{\rm latt}=0.25$.
\item Set $T= 0$ (no \La kicks), $\g= 0.25$, and simultaneously switch to the double well (Higgs) potential. Evolve for $t=3m_s^{-1}$.
\item Evolve with $\g=1.0$ for $t\sim 12m_s^{-1}$ until string loops form.
\item Switch $\g\raw0$, and satisfy condition in eq. \ref{gauss__} (the Gauss constraint) by setting all the momenta to zero.
\item The remainder of the run is now energy-conserving and we evolve this portion with an $\O(\dt^8)$ symplectic integration scheme.
\end{itemize}

\begin{figure*}
  \begin{center}
          \resizebox{150mm}{!}{\includegraphics{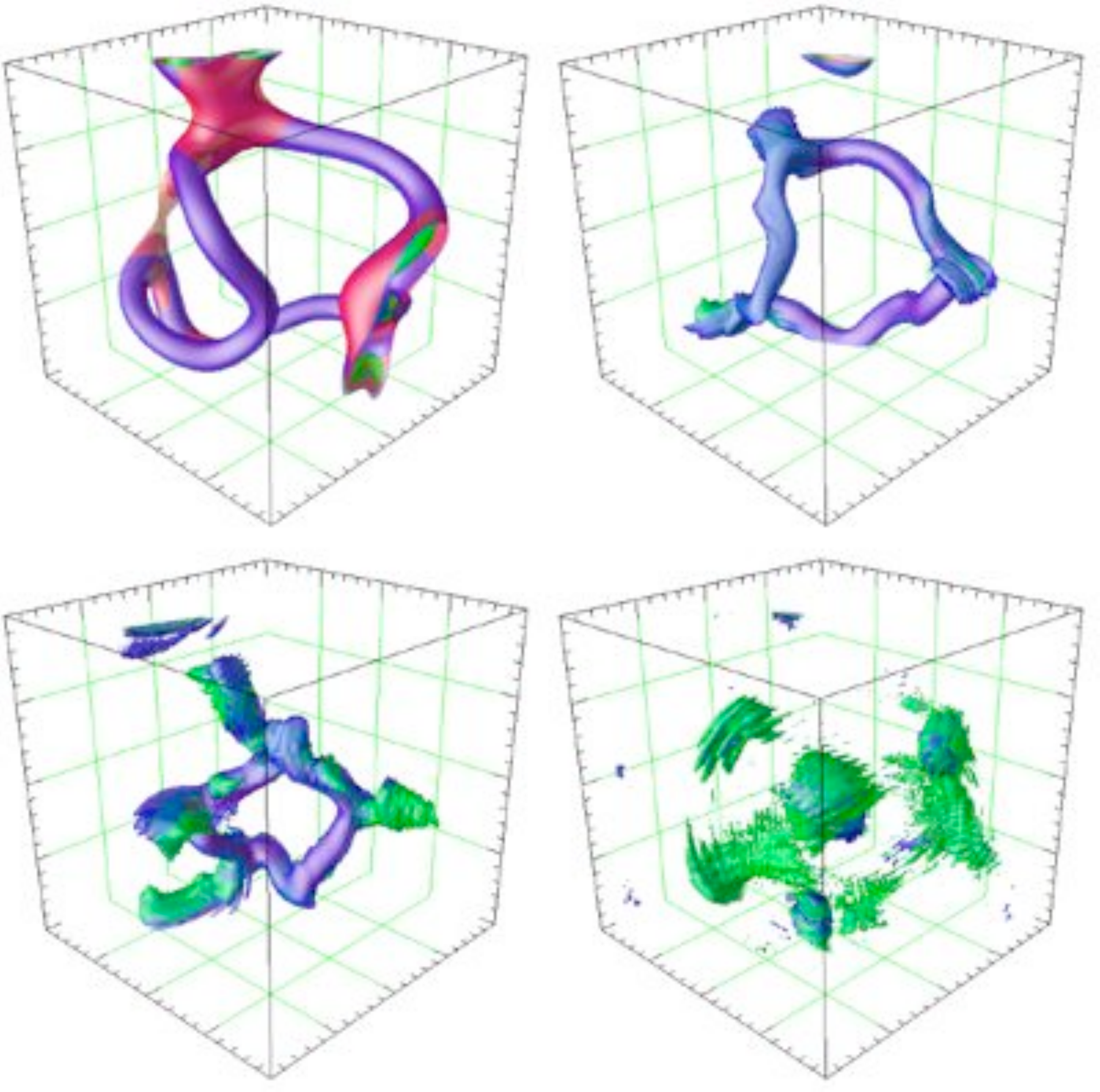}}  
 \caption{(Color online) $4$ snapshots ($t=\lb 12,18,24,43 \rb m_s^{-1} $ increasing left to right) for configuration $C_1$. Plotted is energy density with an isosurface of  $\H_\hx = 0.275$ (blue), $\P^\dagger\P = 0.45$ (red),  $\frac{1}{2}(E^2+B^2) = 0.275$ (green). (Energy isosurface is $25\%$ transparent and when outside of red, appears purple). At the first time-slice the friction has just been turned off. As the string network twists and accelerates, it generates violent changes in the magnetic field at its sharp corners. The shrinking of the toroid forces the flux to get radiated out of these sharp edges, generating the magnetic wakes seen in the last time slice. {\tt \href{http://media.dartmouth.edu/~mgleiser/3dU1g4.mov}{View simulation here.} } 
 } 
    \label{flux_tube_2_osc}
  \end{center}
\end{figure*}

\begin{figure}\centering
\includegraphics[scale=.8]{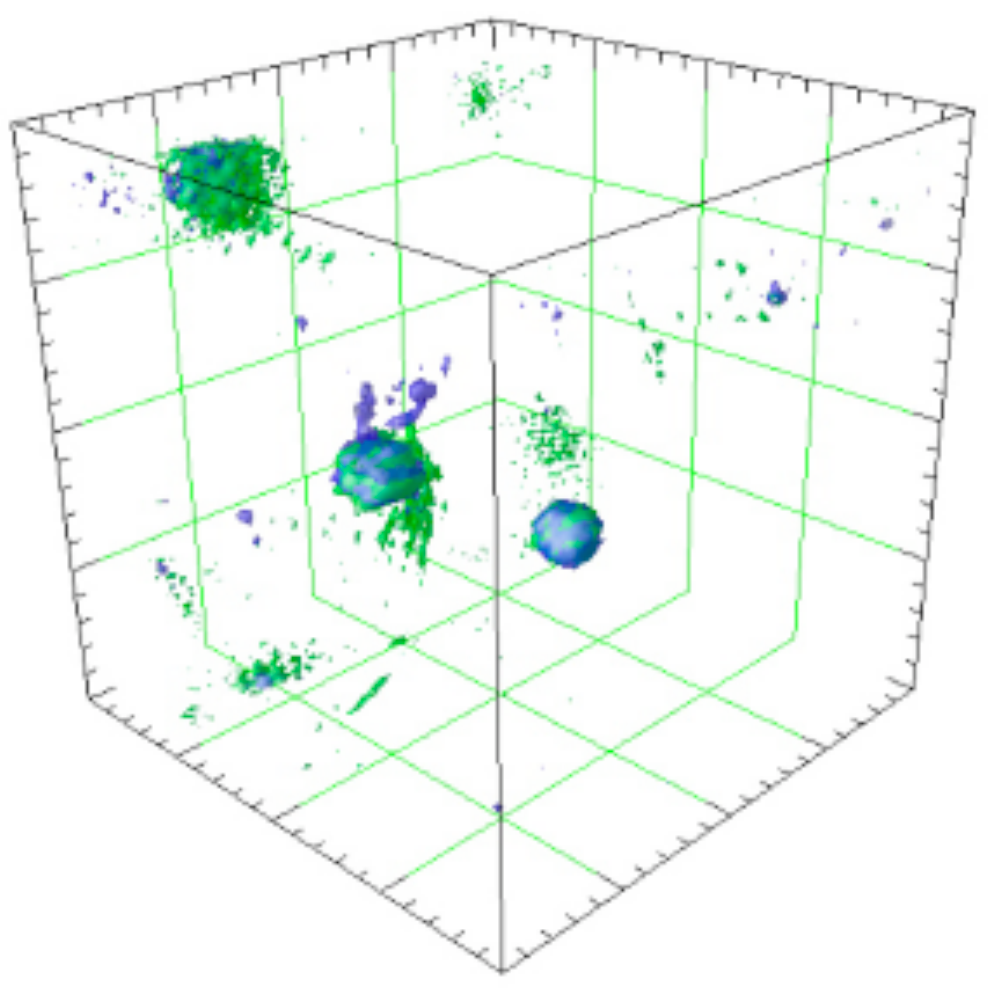}
 \caption{\label{fig:c1:s5} (Color online) Residual oscillon candidates from configuration $C_1$ after the radiation from the flux tube decay has dissipated throughout the lattice.
Time slice at $t=74{m^{-1}_s}$. Energy density isosurface is at $\H_\hx = 0.275$ (blue); scalar field amplitude is at  $\P^\dagger\P = 0.45$ (red not visible since it is inside the energy isosurface); gauge fields energy density is at  $\frac{1}{2}(E^2+B^2) = 0.275$ (green).
 }
\end{figure}

Configuration $C_1$ nicely generates an initial string network. As shown in figure \ref{flux_tube_2_osc}, due to the small lattice size, the strings annihilate quickly into a loop which then violently decays into a few oscillon candidates. Because some of these  candidates are relativistically boosted away from the string annihilation region, they are also Lorentz-contracted in the direction of motion. Some of these decay and we are left with a fairly hot background with three spatially-localized configurations show in figure \ref{fig:c1:s5}. One of these objects still moves quite fast and is noticeably Lorentz-contracted (top left of figure \ref{fig:c1:s5});  it decays after colliding with the larger slow-moving one which has quickly taken a spherical shape (center right of figure \ref{fig:c1:s5}). The second fast structure then decays at $t\sim 200 m_s^{-1}$(not visible in figure), after interactions with the hot background, while the spheroidal object remains until $t\sim 1000m_s^{-1}$. This is our oscillon candidate.

There are a few points to address. First, we call the fast-moving, localized object (on the top left of the figure \ref{fig:c1:s5}) an oscillon candidate and not a radiation wake because we observed that it does not decay according to an expected dispersion relation in the direction of travel or perpendicular to it. One would expect a wake to disperse at least perpendicularly to the direction of travel. The full simulation shows that one can easily distinguish and identify the radiation wakes. {\tt \href{http://media.dartmouth.edu/~mgleiser/3dU1g4.mov}{View simulation here.} }

Second, if these relativistically-boosted structures decay in about $\sim 100-200 m^{-1}$ time units, does it mean they are not very stable? It is hard to say in the context of this simulation. We have seen in $2d$ that even configurations that will evolve to become oscillons, forming at rest and without a hot background (in a near vacuum), can take a long time $\O(10^2 m^{-1}_s)$ to settle down into a coherent object. Thus, if they were boosted with relative velocities to a large amplitude radiation bath at birth or soon after, they may never have settled into an oscillon state. Of course, since this is a Lorentz invariant theory, we can boost any configuration across the vacuum without affecting its stability. But if the boost exposes the object to a relatively fast moving and large amplitude set of radiation modes (so, not near-vacuum), its decay may be catalyzed. Here is an intuitive justification for this fact. Consider a configuration-space attractor characterized by a coherent object. Consider further that it is possible to attach a measure in configuration space, so that there is a well-defined ``distance'' from any configuration to this attractor point.   In the presence of large perturbations, the probability of an initial configuration settling into the attractor will decrease with its ``distance" to it. Back to our simulation, we see that candidate configurations far from reaching the oscillon state may not be stable enough to settle into it in the presence of large perturbations. However, our surviving configuration clearly does.

The third point is that any stability arguments have to take into account stochastic thermal effects. This particular configuration (radiative fields over whole volume plus EFC contribution) has an average energy-density $\bra \H \ket \sim 0.06267$ and an approximate temperature of the same order of magnitude. While the approximate energy density of the oscillon is more than an order of magnitude higher in the core, much of the lower-energy modes which make up the structure will be disrupted by such a temperature. As has been shown in the context of real scalar field oscillons, it is not surprising that thermal noise will compromise the lifetime of coherent states \cite{Gleiser1996}.

Even though no similar stability analysis has been performed in the context of gauged oscillons, temperature effects may be a serious obstacle to the formation of oscillons and to their stability. This is not really that surprising, since we expect symmetries to be restored at high temperatures. The crucial question, then, is if the gauged oscillons can sustain stochastic thermal effects. In other words, if they are present, at what fraction of $T_c$ do they get destroyed?

Back to our simulation, after $\sim 250 m^{-1}_s$ only one oscillon is left on the lattice. It can be treated as a single object with some residual velocity in the midst of a thermal bath. To characterize this object we can look at a few observables. The maximum energy-density is  max$[ \H(t,x,y,z)]  \sim 4$. The fraction carried by the gauge fields is quite large at some times,  $\H_{E^2+B^2}^{max} \sim 3$. Similarly to the $2d$ gauged oscillons, it is clear that the gauge fields play a fundamental role in the oscillon dynamics.  The maximum amplitude of scalar field oscillations is $\P^\dagger \P \sim 0.27$. This is very important: although symmetry is not quite restored at the oscillon core as it is in usual topological defects, it still qualifies as a large nonperturbative coherent fluctuation away from the vacuum state, since the inflection point of the potential is at $\P^\dag\P |_{\rm inf}=0.5$. 
This particular configuration then dies at $t\sim1000{m^{-1}}_s$.

\subsubsection*{Configuration $C_2$}

The details of configuration $C_2$ are: 
\bea 
\lb d, N_L,L,\dx,\dt, g,\l,\eta,T_{\rm latt}  \rb_{C_1} \equiv \\ \nonumber \lb 3,96,19.2,0.2,0.05,5,2,1,0.25 \rb. 
\eea
 We prepare this simulation very similarly to configuration $C_1$:  the gauge coupling is increased from $g=4$ to $g=5$ and we also extend the time in which the dissipation is present for formation of string loop from $t\sim 12 m^{-1}_s$ to $t\sim 26m^{-1}_s$.

\begin{itemize} \label{recipe_c_2}
\item Set all fields to their vacuum configurations in the quadratic potential $\P^\dagger \P$.
\item Thermalize using \La dynamics with $\g=1$ for $t\sim 10 m^{-1}$ at $T_{\rm latt}=0.25$.
\item Switch $T\raw 0$ and $\g\raw 0.25$ and simultaneously switch to the double well (Higgs) potential. Evolve for $t=3m^{-1}_s$.
\item Evolve with $\g=1.0$ for $t\sim 26m^{-1}_s$ until string loops form.
\item Set $\g=0$, and satisfy condition \ref{gauss__} by setting all the momenta to zero
\item The remainder of the run is now energy-conserving and we evolve this portion with an $\O(\dt^8)$ symplectic integration scheme.
\end{itemize}

\begin{figure*}
    \centering
\resizebox{180mm}{!}{\includegraphics{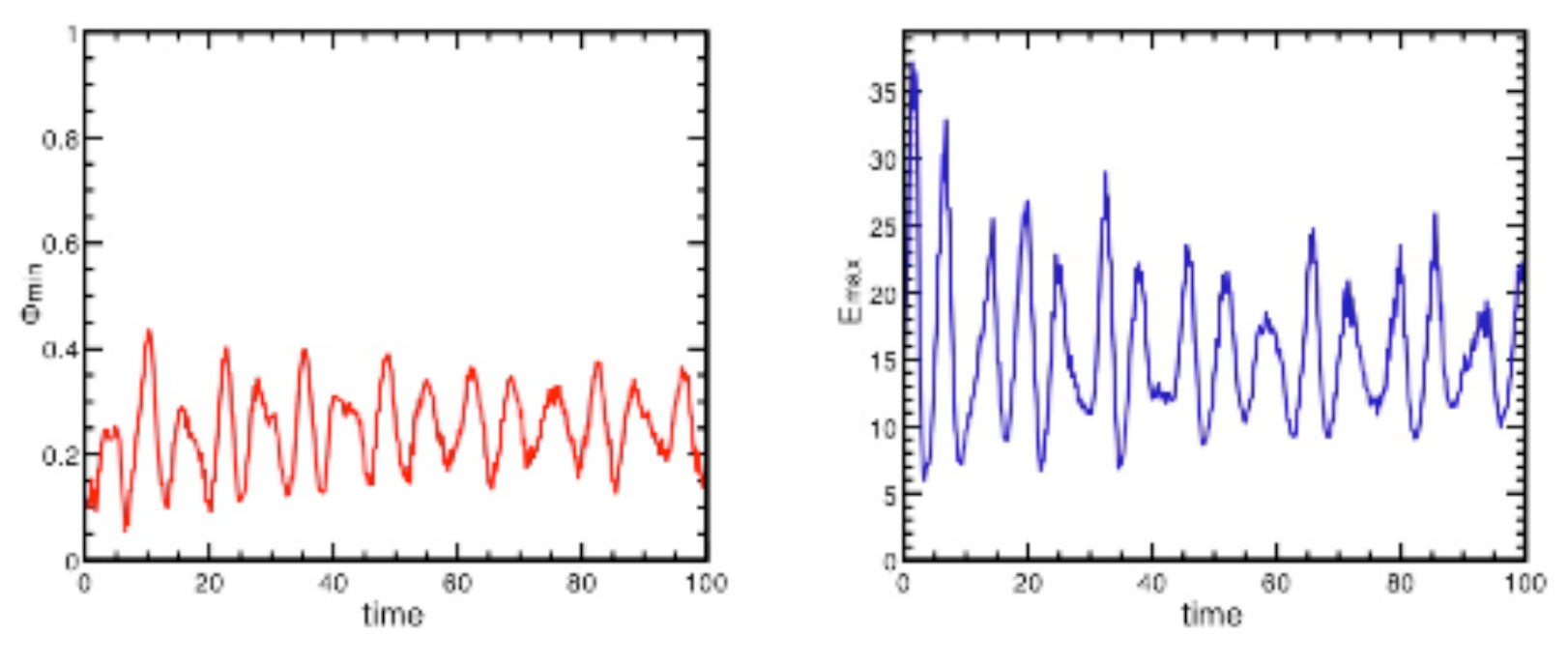}}
\caption {(Color online) On the left is plotted the minimum value of $\P^\dag\P$ for the $\b=0.04$ oscillon configuration in figure \ref{osc_max_min_slice}. On the right is the maximum of the energy density $\H(t,x,y,z)$. }
\label{osc_max_min1}
\end{figure*}

An oscillon is formed from this configuration and we observed it for a time of $ t \sim 15000 m^{-1}_s$. At this point we stopped the simulation as the oscillon had not changed for thousands of time units, continuing to drift across the lattice. In figure \ref{osc_max_min1} we plot for a short time the evolution of two important observables related to this oscillon: maximum of energy-density max[$\H$] and minimum amplitude of scalar field ${\rm min}[\P^{\dagger}\P]$. Compare with similar plot for equivalent oscillon in $2d$ in figure \ref{fig:2d_oscil}. Cleaner data will be obtained below.

Now that we know that these objects exist in $3d$, we can move on to a more in-depth study of their properties. The problem with the two previous configurations is that they leave too much energy on the lattice, making it hard to study the oscillon accurately. This extra energy also catalyzes their decay. Next, we propose an approach to remove all the spurious energy so that we can investigate $U(1)$ oscillons in more detail.

\subsection{Isolating the $3d$ $U(1)$ Solution}

\begin{figure} \centering
\includegraphics[scale=.6,angle=0]{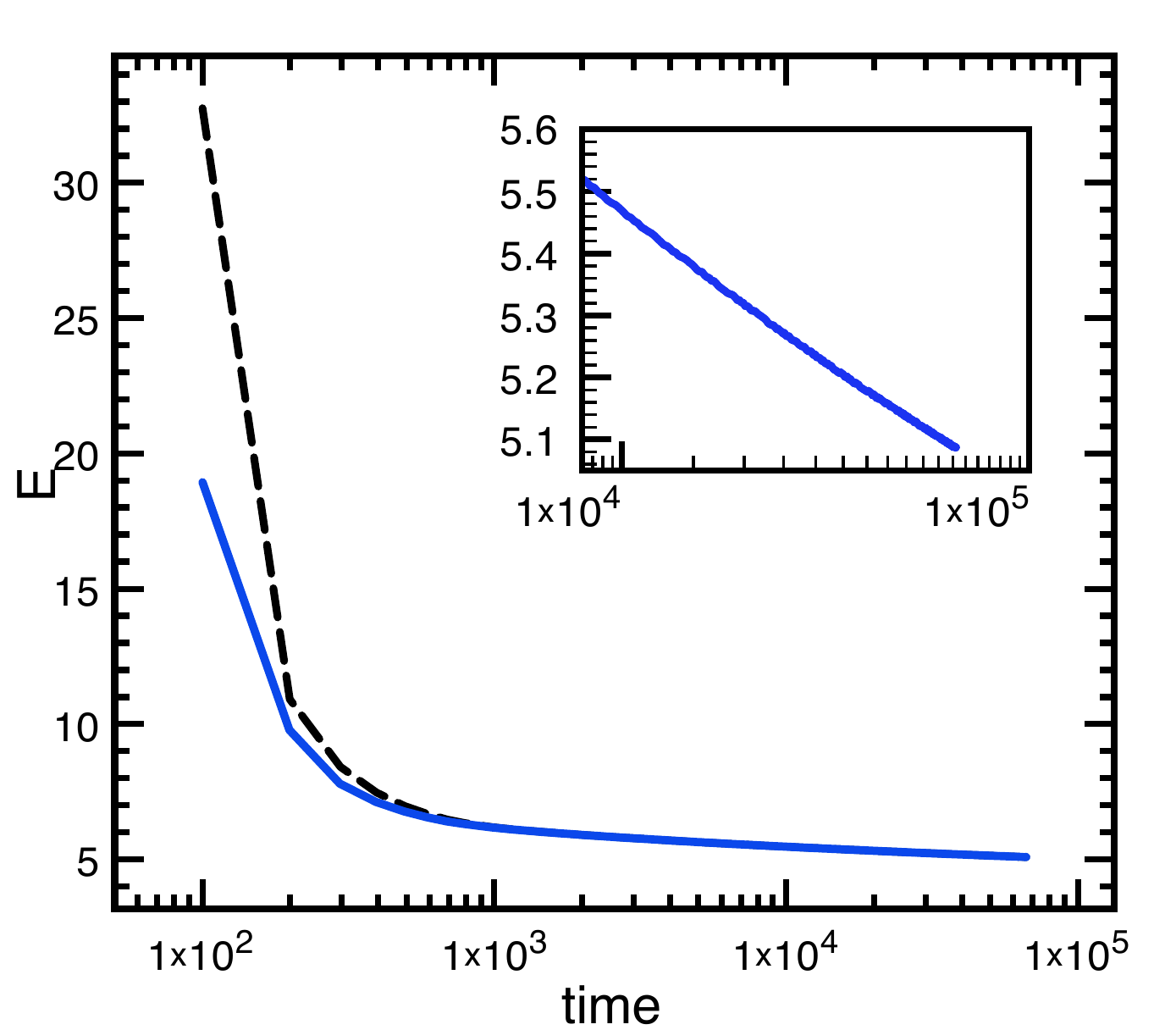}
 \caption{\label{fig:radiation}(Color online) Time evolution of the total energy (dashed) and of the energy within a sphere of $R=4$ for a $\beta =0.01$ oscillon. After a quick shedding of a large amount of energy, the onset of the oscillon stage at at time $t \sim 10^3$ is characterized by a very slow energy decay accurately fitted by $\dot{E}(t)=-c_1 t^{-c_2}$, where $c_2\simeq 1.2$. The inset shows the numerical value of the energy and the fitted curve (indistinguishable) for $t \geq 10^4$.  } 
\end{figure}

\begin{figure*}
  \begin{center}
\resizebox{150mm}{!}{\includegraphics{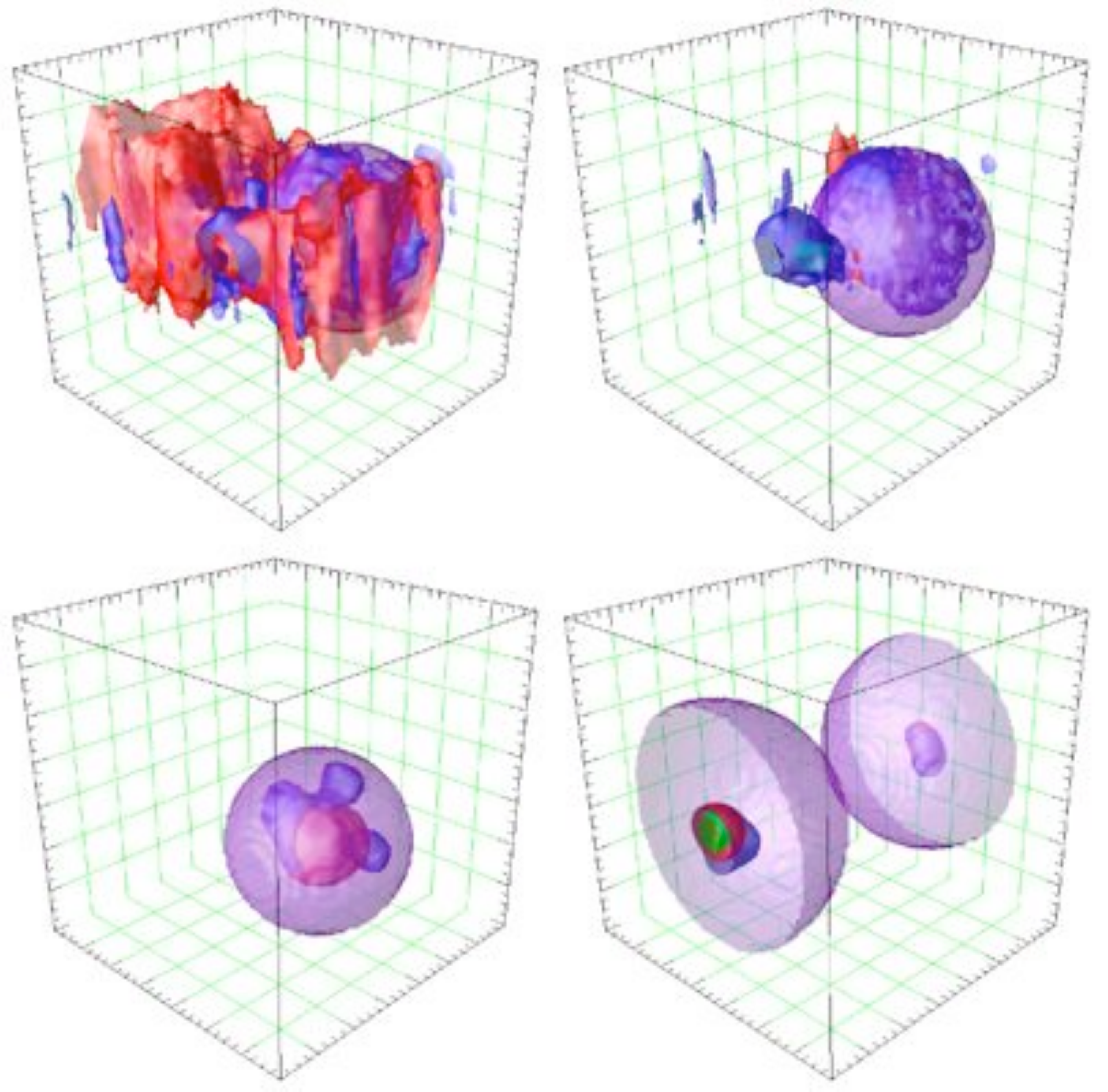}}
    \caption{(Color online) $4$ snapshots ($t=\lb 0,1,16,90 \rb m_s^{-1} $) of the initial stages of a long-lived $3d$ $U(1)$ oscillon. Plotted is total energy density (blue) at an isosurface of $0.05$. Energy density isosurface in the gauge fields (green) is also at $0.05$, while  for the scalar field magnitude $\P^\dag\P=0.9$ (red). Size of plotted area is $1906 m^{-3}_s$ and $\b=0.04$. The energy of the resulting oscillon is about $E_{osc}\sim9$. The spherical shell (magenta) in lower plots denotes the friction wall which follows the oscillon throughout the lattice. Because of the toroidal boundary conditions, the nearly spherical profile of the low energy outer surface of the oscillon is visible as a cross section at $t=90$ (lower right). Note also that a second oscillon candidate is present at $t=1$ (top right), but doesn't survive long due to the friction. {\tt \href{http://media.dartmouth.edu/~mgleiser/3d_q5_osc_F.mov}{View simulation here.} } }
    \label{osc_max_min_slice}
  \end{center}
\end{figure*}

In order to isolate the oscillon, we take an initial configuration that we know allows for at least one from flux-tube annihilation, that is, for the annihilation of two long strings with opposite magnetic fluxes inside, the equivalent of $2d$ vortex-antivortex annihilation. After the flux tubes interact and annihilate, we search the lattice for the maximum value of the energy density, ${\rm max}[{\cal H}(t,x,y,z)]$ which hopefully correlates with the presence of an oscillon. After finding the maximum and tracking it for a while to make sure it is sufficiently long-lived, we place a spherical friction wall at a radius $r\geq R_f$ from it (the choice of $R_f$ to be made explicit soon) to dissipate all modes outside the EFC, effectively ``mopping'' the lattice from possible destructive perturbations that may affect the formation and longevity of the oscillon. The fields within the cavity at $r< R_f$ settle into an oscillon, which moves about the lattice carrying the friction wall along. When the energy for $r\geq R_f$ is only very slowly decreasing we know that we have dissipated most of the excess energy in the lattice volume; any residual energy within the cavity is contributed by the oscillon itself.  Once the average energy density outside the cavity has decreased to $\sim\O(10^{-4})$ times the oscillon's energy (that is, the energy for $r<R_f$),  we can turn the friction off and watch the oscillon drift about the lattice. Using this method, we have seen oscillons live for longer than $t \geq 69000m^{-1}_s$ for $\b=0.04$ without showing any signs of instability. In figure \ref{fig:radiation} we show the oscillon energy as a function of time for a simulation with $\b=0.01$, where its long time behavior can be explicitly seen. Note that even though our phase diagram stops at $\b=0.04$ this doesn't mean that there are no oscillons formed for smaller $\b$. This why we include the example in the figure with $\b=0.01$. The small $\b$ analysis is incomplete due to CPU limitations and not physics, as is the case for $\b > \b_c$. 
This method also allows the approximate determination of the oscillon's energy, obtained by a volume integral over the lattice [$E_{\rm osc}(\b=0.04)\sim9$], and the value of the energy at the oscillon's core, which is the tracked value ${\rm max}[{\cal H}(t,x,y,z)]$. This value and the value of the minimum of $\P^\dag\P$ are plotted in figure \ref{osc_max_min1} for a short time after the oscillon has formed and the friction wall had been turned off (marking the $t=0$ in the plots.) These plots correspond to the isosurface plots in figure \ref{osc_max_min_slice}.

More specifically, the friction is implemented by constructing a shell around the location of max$[\H(t,x,y,z)]$ which vanishes for a radius $r \leq R_f = 4.0$ and equals $\g=\tanh(r-R_f)/2$ for $r\geq R_f$. The oscillon is safely localized within this shell. Although the implementation of the friction wall does a great job isolating the oscillon solution, it obviously violates the Gaussian constraint \ref{gauss__}, as one can see by rederiving current conservation from the equations of motion with this nontrivial form of $\g$. However, the fact that what violates Gauss's constraint is a viscosity term only acting in a region with negligible energy-density contribution from the configuration of interest is helpful. If the violations are far enough from the region of interest -- that is, for $r < R_f$ -- then the simulation is trustworthy since the violations continue to decrease in magnitude as $r$ increases. We also note that usually, when there are numerical instabilities or violations in constraint equations, these will tend to create exponential instabilities. This is not the case here, as our numerical method is perfectly stable in the presence of these nonpropagating charge densities.

\begin{figure*}
  \begin{center}
\resizebox{150mm}{!}{\includegraphics{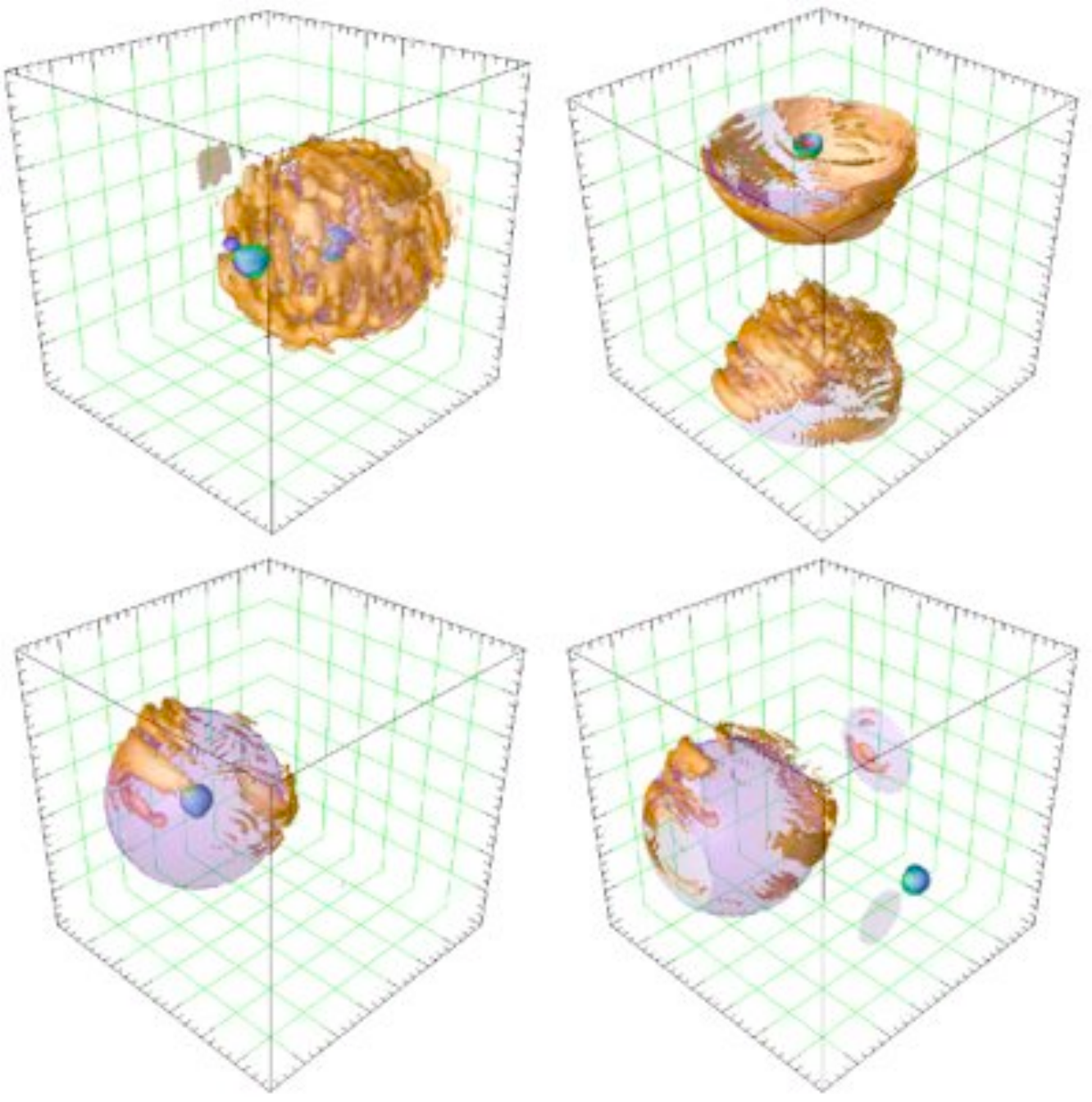}}
   \caption{(Color online) $4$ snapshots ($t=\lb 1,45,192,520 \rb m_s^{-1} $) of $3d$ $U(1)$ oscillon displaying violation of Gaussian constraint. Plotted is energy density (blue) with an isosurface of $1.00$ marked for scale. Energy density in the gauge fields (green) is also marked at $0.10$ while $\P^\dag\P=0.4$ (red). The absolute value of the violation of the Gaussian constraint is shown in orange at an isosurface of $5\times 10^{-3}$. Size of plotted area is $1906 m^{-3}_s$. This is also a $\b=0.04$ configuration. The energy of the resulting oscillon is about $E_{osc}\sim9$. The spherical shell (purple, bottom plots) denotes the friction wall that follows the oscillon throughout the lattice. After $t\geq 200m^{-1}_s$ we turn the friction wall off, but plot its last position to highlight that the region of Gaussian violation stays anchored in space.  {\tt \href{http://media.dartmouth.edu/~mgleiser/3dU1gauss.mov}{View simulation here.} }
   }       
    \label{fig:gauss_vio_slice}
  \end{center}
\end{figure*}

To investigate the violation in Gauss's constraint we take the absolute value of equation \ref{gauss__}. This we then plot along with the friction wall and the localized structure which has formed inside of it, the oscillon. Hopefully, from this experiment we can gain some understanding of what effect, if any, the Gaussian constraint violation has on the dynamics. In figure \ref{fig:gauss_vio_slice} we show the oscillon (in blue an green) up to $t\sim 500m^{-1}_s$ as it moves through regions of Gaussian violation (in orange) at about the $10^{-3}$ level without any discernible variation of its observable max$[\H]$ and min$|\P^\dag\P|$ oscillations as compared to oscillons formed in systems with Gaussian violations at the $10^{-12}$ level of absolute magnitude. As the friction wall is turned off at $t\sim 200m^{-1}_s$ (bottom left), the extra charge density from the Gaussian violation stays fixes in space, while the oscillon keeps moving about the lattice with no discernible effect (bottom right). {\tt \href{http://media.dartmouth.edu/~mgleiser/3dU1gauss.mov}{View simulation here.} }

\begin{figure} \centering
\includegraphics[scale=.6,angle=0]{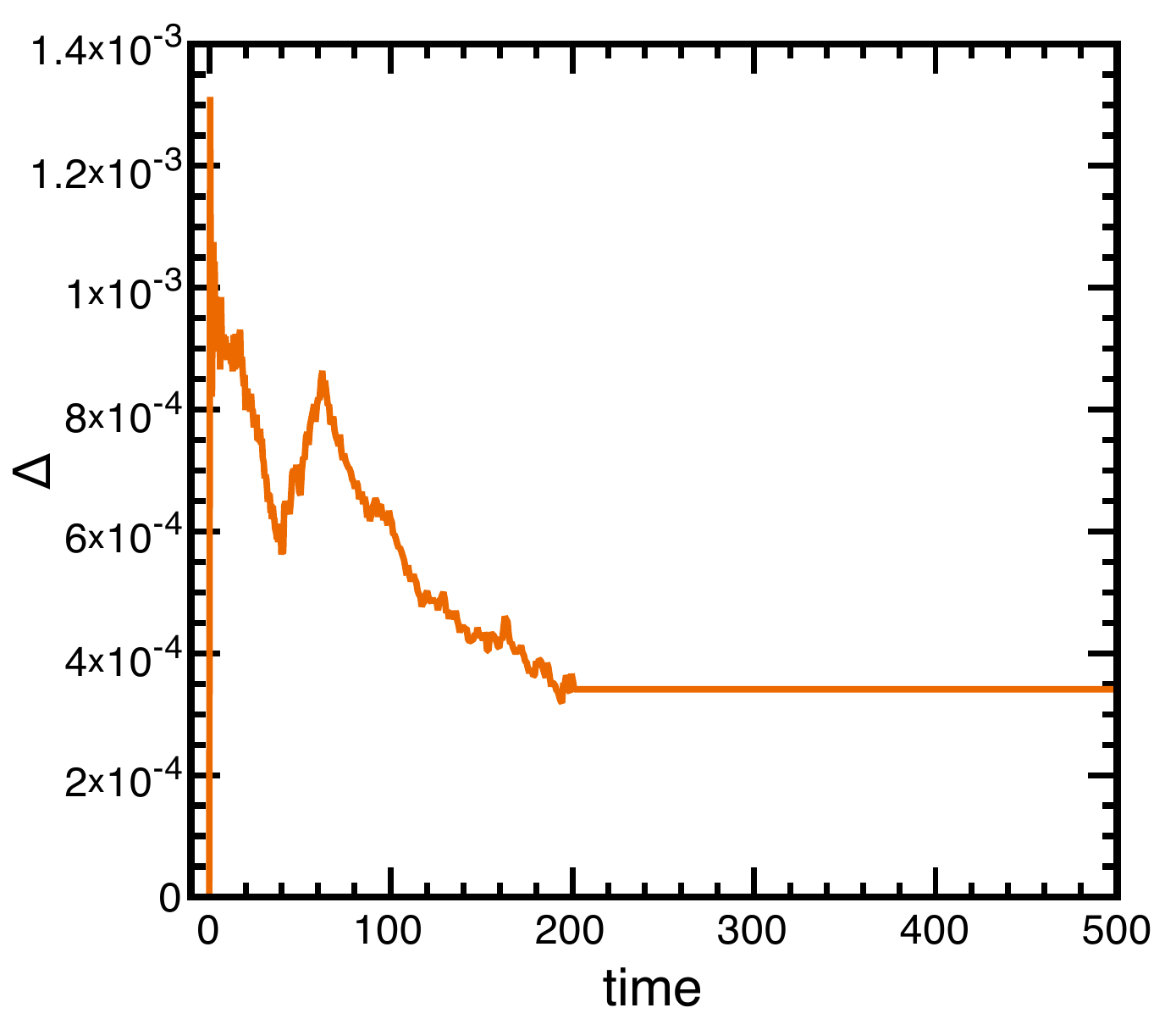}
\caption{(Color online) Global expectation of the violation $\Delta(t)=\bra |{\frac{1}{\dx} \sum_{i} ( E^i_\hx - E^i_{\hx -\hat{i}} ) - 2 q \im (\Pi_x^\dagger \P)}| \ket$ for the simulation in figure \ref{fig:gauss_vio_slice}. For simulations without a spatiotemporally complex support function for friction, $\Delta(t) \sim 10^{-12} (a t)^{b}$, where $a$ and $b$ are small numbers.   }
\label{fig:gauss_vio_av}  
\end{figure}

In figure \ref{fig:gauss_vio_av} we plot the expectation value of the violation of Gauss's constraint $\Delta(t)=\bra |{\frac{1}{\dx} \sum_{i} ( E^i_\hx - E^i_{\hx -\hat{i}} ) - 2 g \im (\Pi_x^\dagger \P)}| \ket$ as as function of simulation time. At $200m^{-1}_s$ we turn off the spherical wall of friction and then any violation left on the lattice is frozen due to conservation of charge to our numerical precision.

Based on these observations, we did not see any undesirable effect due to the presence of small amounts of violations in $\Delta(x,t)$. What we gain from this relaxation of numerical rigor is a path to investigate the EFC for a long time.

\section{Constructing a $3d$ $U(1)$ Phase Diagram in Configuration Space} \label{section:phasediagram}

Once we have numerically shown the existence of a  time-dependent EFC in a given theory, the next step is to categorize the region of parameter space in which this solution is an attractor configuration. From our experience, it is almost impossible in some regions to actually find an object, despite hints that it might exist. So, the method we chose to predict where the object lives in configuration space is to pick an observable which can be used to characterize certain properties of EFCs in a region where they are known to exist, and then study that observable as a function of parameter space. With this we can predict by extrapolation the regions with no EFCs. As in $2d$, the results can be usefully organized in a phase diagram. 

The observables we choose must vanish when no oscillons are present. Since, when oscillons are present, the maximum local energy-density on the lattice max$[{\cal H}(t,x,y,z)]$ is much larger than the background noise, and because we are searching for large amplitude structures, we will use the maximum energy-density and the minimum min[$\P^\dag\P$] amplitude as order parameters.

\begin{figure} \centering
\includegraphics[scale=.6,angle=0]{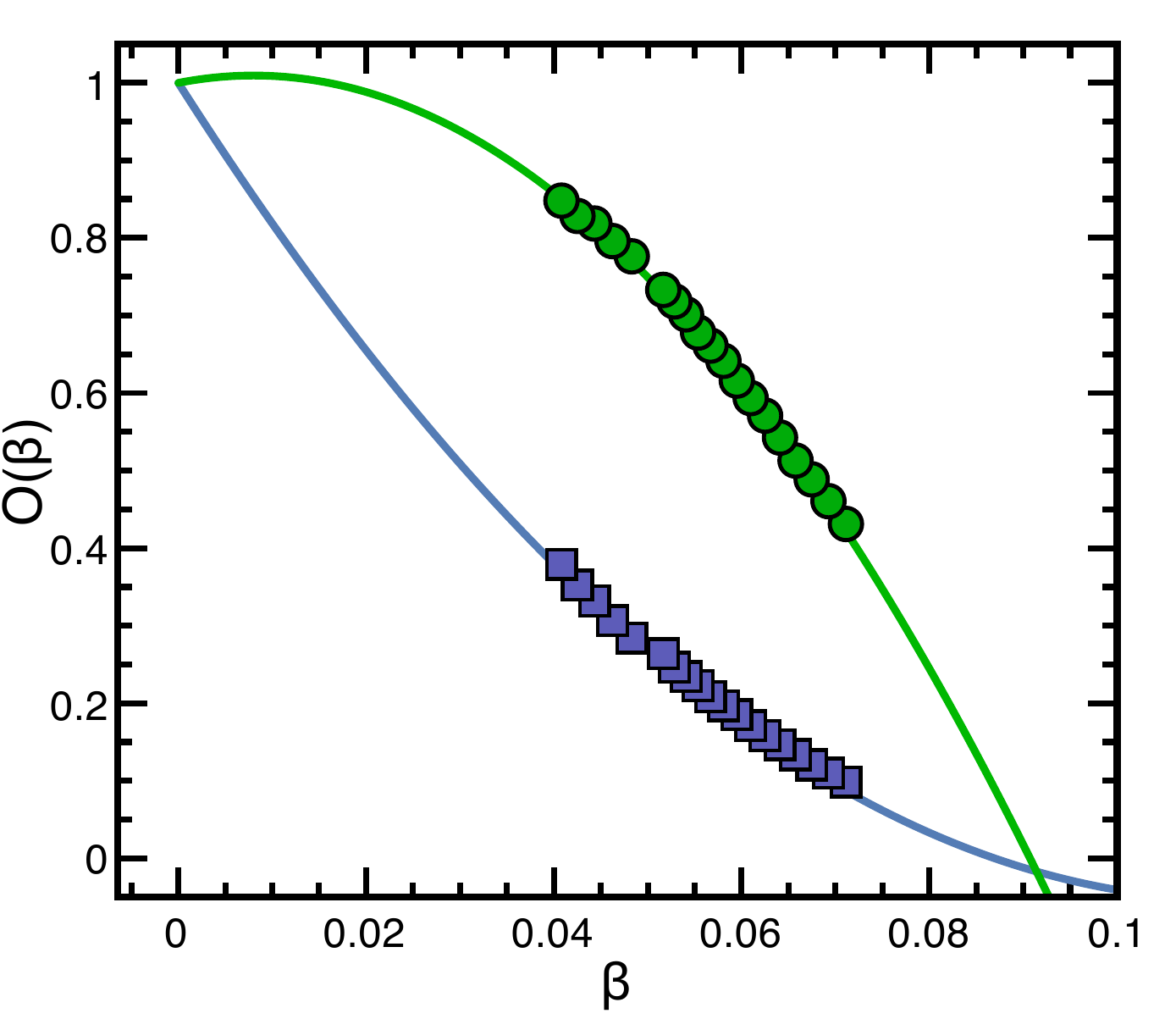}
 \caption{\label{phasediag_3d} (Color online) Phase diagram for the $3d~U(1)$ flux-antiflux tube annihilation $\rightarrow$ oscillon transitions. Plotted as a function of $\b$ are the order parameters $O_{E}(\b)\equiv E_{\rm max}(\beta)/E_{\rm max}(0)$ with numerical values in boxes,  and $O_\P(\b)\equiv (1- \P^\dag\P_{\rm min}(\b))/(1- \P^\dag\P_{\rm min}(0))$ in circles. The continuous lines are fits which we use to extrapolate to the critical value where no oscillons are expected to form by this mechanism. We find $\beta_c\sim 0.0893$ and $\beta_c\sim 0.0908$ for the  $O_{E}(\b)$ and $O_\P(\b)$ order parameters, respectively.}
\end{figure}

Our basic method will be to form an oscillon following similar steps to configuration $C_2$ in section \ref{recipe_c_2}, except that we will repeat the procedure for various $\b$. On the lattice, we track ${\rm max}[\H]$ and ${\rm min}[\P^\dag\P]$ as a function of time, and take a running average of these oscillating observables.  As they converge to a near-constant value (usually at times $t\geq 10^{3}m^{-1}_s$), we plot them in a phase diagram as shown in figure \ref{phasediag_3d}. We can then use the diagram to extrapolate and find the critical parameter $\b_c$ beyond which oscillons do not form. The basic method can be summarized as follows: 
\begin{itemize}
\item Prepare a configuration similarly to $C_2$ in section \ref{recipe_c_2}. If an EFC is formed (confirmed by identifying a persistent local maximum in energy density), then measure ${\rm max}[\H]$ and ${\rm min}[\P^\dag\P]$  to confirm it is an oscillon. If not, repeat with a different Langevin realization using the same initial temperature.
\item If an oscillon forms then introduce the moving friction wall until the energy-density outside the coherent object is negligible compared to ${\rm max}[\H]$. ($t\sim 250m^{-1}_s$ has proven to be sufficient.)
\item Turn off the friction and allow the system to evolve using an $\O(\dt^8)$ symplectic integration routine. 
\item Track ${\rm max}[\H]$ and ${\rm min}[\P^\dag\P]$ for $2500m^{-1}_s$.
\item Construct a running average of these quantities calling them $E_{\rm max}$ and $\P^\dag\P_{\rm min}$, respectively. If this running average converges to a near-constant value, then take this value as a point in the phase diagram.
\end{itemize}

It is essential that within a reasonable amount of time the time-averaged observables show only negligible time variation. If we cannot (after some reasonable amount of attempts) find an oscillon, we move to a lower $\b$ since our intuition from the $2d$ work indicates this favors their production. Fortunately, there is a regime in which the observables converge well in a short time, allowing us to fit a function and extrapolate to the critical point in $\b$ above which no oscillon should form.

 \begin{figure*}
    \centering
      \resizebox{160mm}{!}{\includegraphics{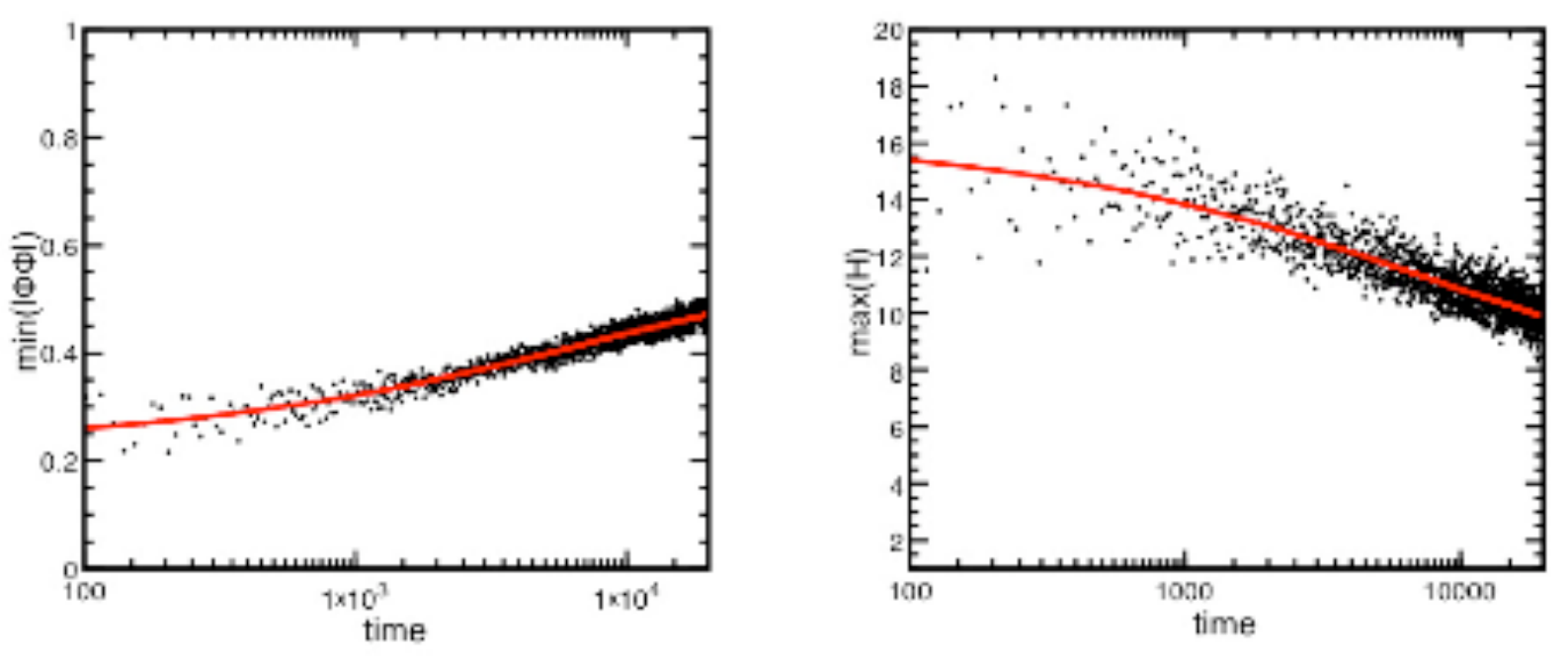}}
\caption {(Color online) On the left is plotted the minimum value of $\P^\dag\P$ for an $\b=0.04$ oscillon configuration. The right is the maximum energy density $\H$. These correspond to $t=2\times 10^{4}m^{-1}_s$ in data and highlight the very slow leaking of energy from oscillons. When ${\rm min}[\P^\dag\P] \geq 1/\sqrt{2}$ we expect the oscillon to decay completely.  The red lines are $3$-parameter fits to the data which are identified in equations \ref{param_emax} and \ref{param_phi}. }
\label{osc_max_min}
\end{figure*}

In figure \ref{phasediag_3d}, we present the averaged order parameters 
\be 
O_E(\b)\equiv  \frac{E_{\rm max}(\beta)}{E_{\rm max}(0)}, 
\ee 
where $E_{\rm max}(\beta)$ (blue curve) is the value take from the running average of ${\rm max }[\H]$, and in the green curve we plot 
\be 
O_\P(\b)\equiv \frac{1- \P^\dag\P_{\rm min}(\b)}{1- \P^\dag\P_{\rm min}(0)}.
\ee 
The numerical values are fitted from the region where we trust the data and then extrapolated to an effective $\b_c$. We find $\b^{E}_c\sim 0.0893$ and $\beta^\P_c\sim 0.0908$ for the energy and $\P$ order parameters, respectively. These are higher than the region in which we actually see the oscillons form. 

Take, for example, $\b^\P_c$ obtained from the fit to $\P^{\dagger}\P$. We extrapolated to when the ${\rm min}[\P^\dag\P]\rightarrow 0$, thus matching the vacuum. This is not restrictive enough. Generally, these oscillons decay if their core oscillations do not probe near the inflection point of the potential. If we had extrapolated to the inflection point $\P\sim 1/\sqrt{2}$,  $\b_c$ would change to $\b_c\sim 0.078$. This more restrictive criterion moves $\b_c$ to almost exactly where we stop seeing oscillons from flux tube decay.

In figure \ref{osc_max_min}, the time evolution of these (unscaled) observables is presented for $\b=0.04$. This configuration is at about the limit of what we can resolve numerically due to the extended time it takes to settle into an object with a well-defined maximum and minimum for the observables. It is possible to define a $3$ parameter fit to the data, 
\be 
\bra {\rm max}(\H)\ket\sim c_0 + c_1 e^{-c_2 \sqrt{t}},  
\label{param_emax} 
\ee 
while 
\be 
\bra {\rm min}(\P^\dag\P)\ket\sim c_0 + c_1 (1- e^{-c_2 \sqrt{t}}). 
\label{param_phi} 
\ee 
This parameterization can be useful in identifying the asymptotic value of the order parameter early in the simulation, as we know that $\bra\O \ket (t\raw\infty) \sim c_0$.  It would be instructive, but beyond the scope of this work, to recreate figure \ref{phasediag_3d} for a wider range of $\b$ to test if our extrapolation to $\b_c$ indeed works as we expect it to.

In this section we have shown examples of oscillons forming in various configurations from flux-tube decay in the $3d~U(1)$ theory. We have also defined order parameters to identify their presence and properties, and mapped out the region where we observed them numerically. Using these results, we have predicted the range of $\b$ where we can expect to see oscillons. For $\b < 0.04$ we have not found oscillons but predict their existence. It is a challenging numerical problem, as we have two very different scales in this regime (the scalar and vector masses). 

\section{Formation Mechanisms and Fine Structure}


In the previous sections we described how $U(1)$ oscillons
form directly from flux-antiflux  tube interactions with no residual topology. On large lattices, there is a high probability of forming a complex flux-tube network which will include the needed precursor loops to form such oscillons.

\begin{figure*}
  \begin{center}
    \resizebox{150mm}{!}{\includegraphics{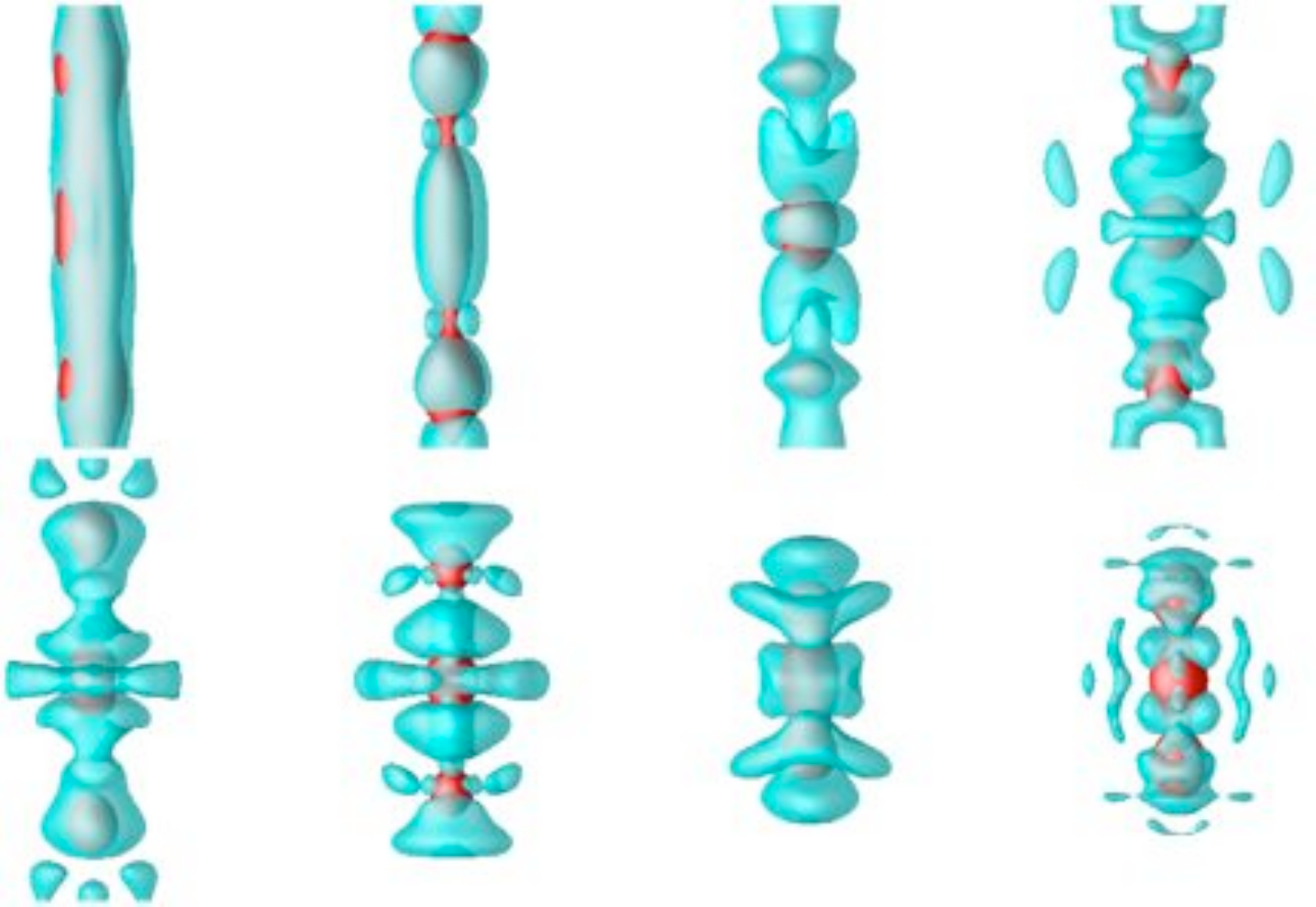}}
    \caption{(Color online) $8$ snapshots ($t=\lb 78,79,81,83,84,90,94,106 \rb m^{-1} $ increasing left to right) for a $\beta=0.01$ oscillon. Plotted is a low energy electric energy-density isosurface of  $\frac{1}{2}E_i \cdot E_i = 0.03$ (cyan) and the condensate density $\P^\dagger\P = 0.5$ (red). Each slice represents an area of $12.5 \times 9 m^{-2}_s$.}
    \label{fig:backbone}
  \end{center}
\end{figure*}

\begin{figure*} \centering
  \begin{center}
  \resizebox{150mm}{!}{\includegraphics{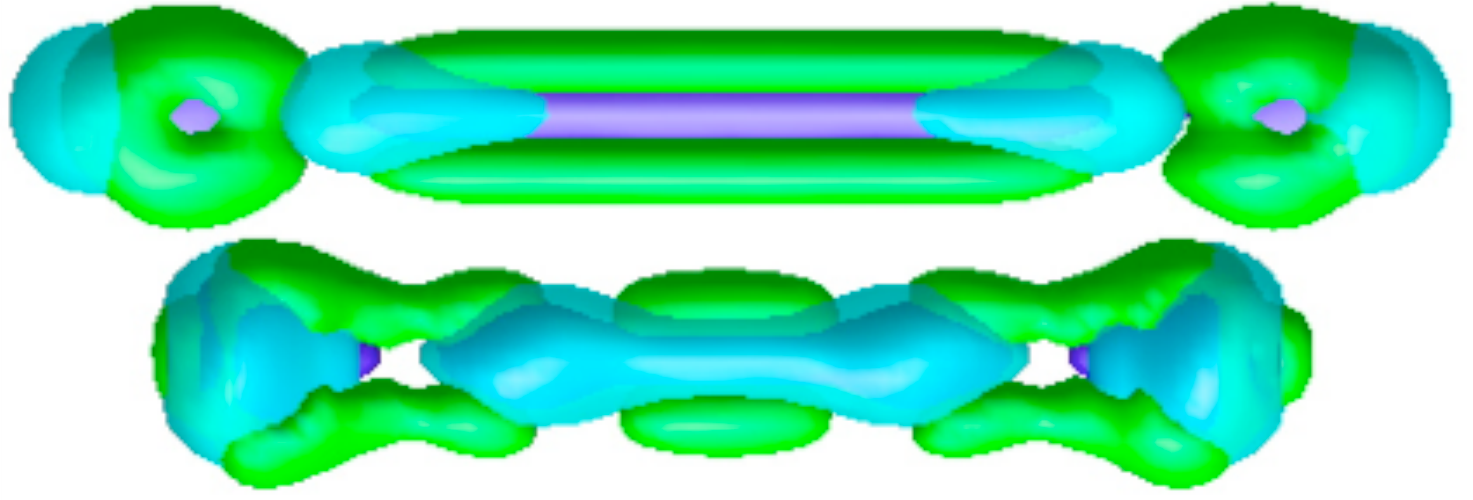}}
    \caption{\label{fig:birth}  (Color online)
    Formation dynamics of an elongated $\hat{z}$ symmetric $3d$ oscillon. $\H(x,y,z)=1.6$ is represented by the dark blue isosurface. Electric energy-density $E^2=0.2$ (cyan), and magnetic energy-density $B^2=0.1$ (green). The bottom configuration is $11m^{-1}_s$ forward in time from the top. The plots represent $\sim 11m^{-1}_s$ across the longest direction. Note that the energy-density tube between the two oscillons in the extremities oscillates between being dominated by its electric and magnetic contributions.} 
      \end{center}
\end{figure*}

\begin{figure*}
  \begin{center}
   \resizebox{150mm}{!}{\includegraphics{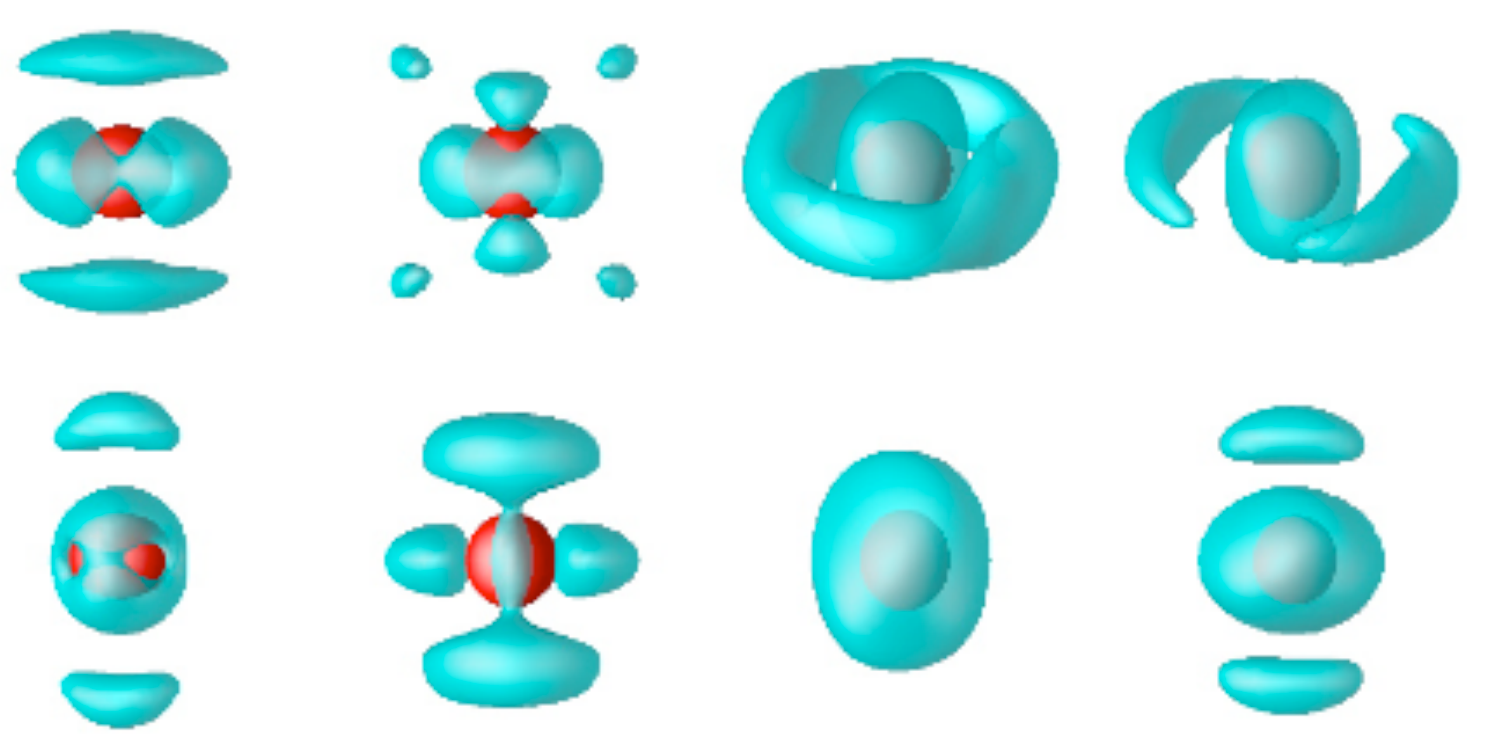}}
    \caption{(Color online) $8$ snapshots ($t=\lb 129,143,155,156,158,164,166,202 \rb m^{-1}_s $ increasing left to right) for a $\beta=0.01$ oscillon. Plotted is a low energy electric energy-density isosurface of  $\frac{1}{2}E_i \cdot E_i = 0.015$ (cyan) and the scalar condensate amplitude $\P^\dagger\P = 0.5$ (red). Each slice represents an area of $25m^{-2}_s$.  The radius of the scalar condensate isosurface is approximately $r\sim 0.75 m^{-1}$.}
    \label{electric_sphere_harmonics}
  \end{center}
\end{figure*}

\begin{figure*}
  \begin{center}
     \resizebox{150mm}{!}{\includegraphics{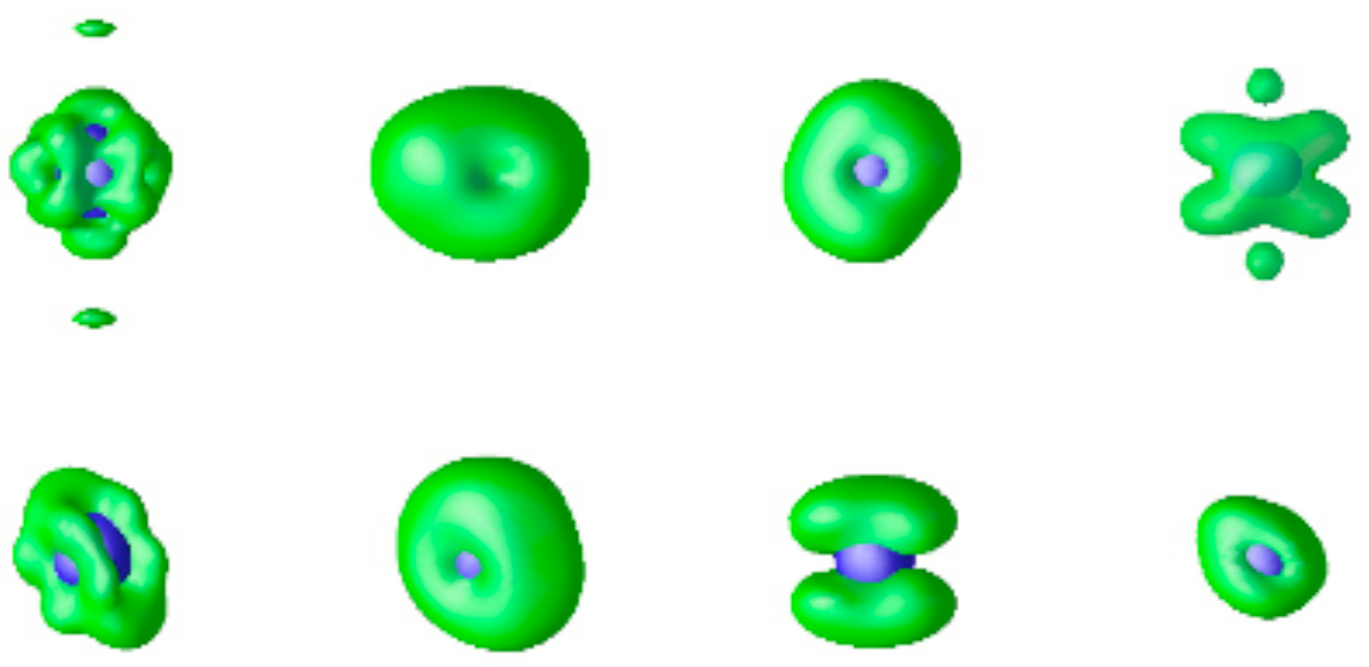}}
    \caption{(Color online) $8$ snapshots ($t=\lb 133,155,166,172,178,185,202 \rb m^{-1}_s $ increasing left to right) for a $\beta=0.01$ oscillon. Plotted is a low energy magnetic energy-density isosurface of  $\frac{1}{2}B_i \cdot B_i = 0.0125$ (green) and the total energy density $\H = 2.5$ (blue). Each slice represents an area of $25m^{-2}_s$. The radius of the energy density isosurface is approximately $r\sim 0.75 m^{-1}_s$.}
    \label{magnetic_sphere_harmonics}
  \end{center}
\end{figure*}

\begin{figure} \centering
\includegraphics[scale=.45,angle=90]{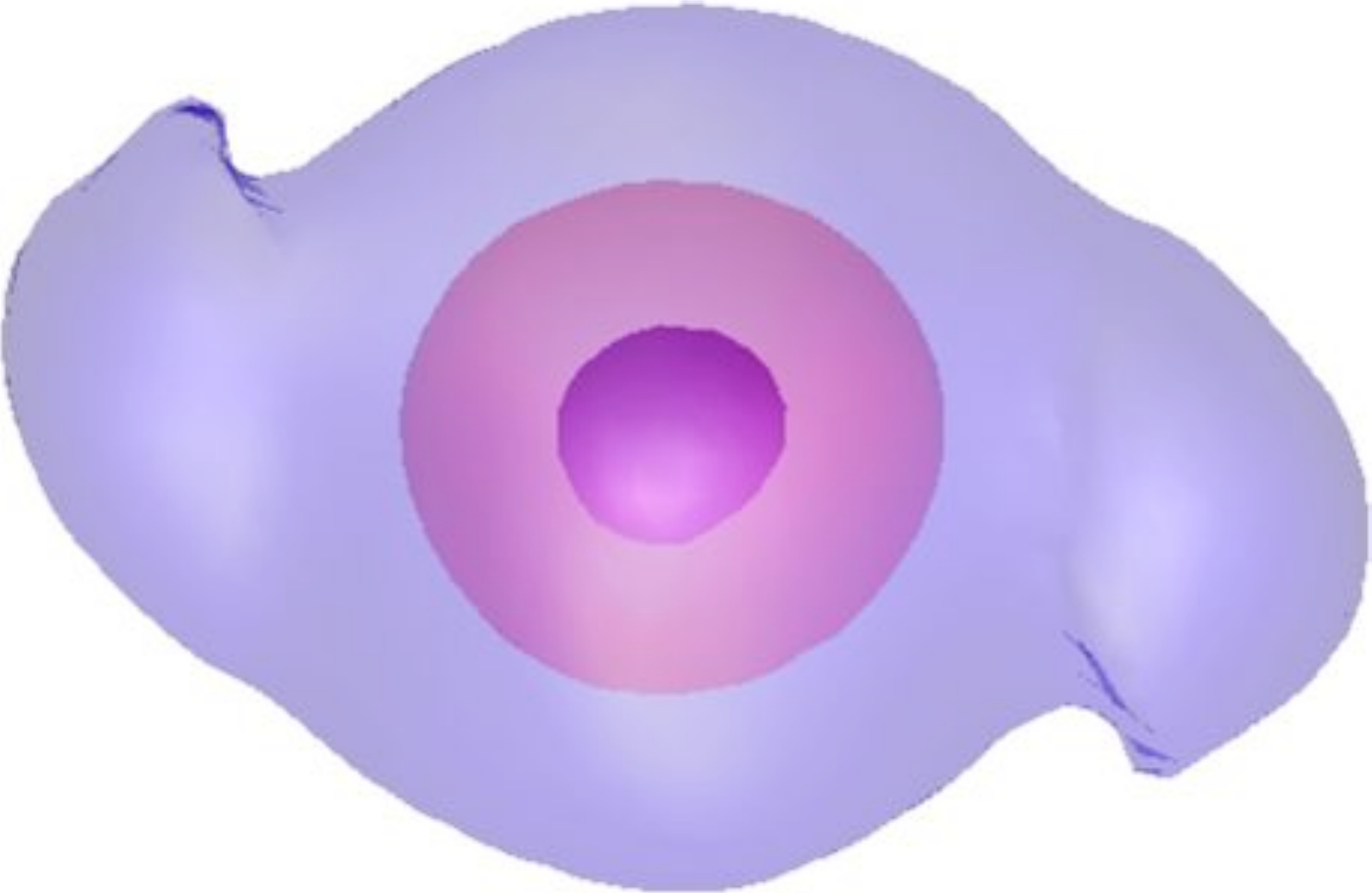}
\caption{\label{energy_various_isosurfaces} (Color online)  Energy density isosurfaces for a $\b=0.01$ oscillon at $t\sim1600m^{-1}_s$ after formation. Isosurfaces shown are at $\H=\{0.01({\rm blue}),0.25({\rm magenta}),5.0({\rm violet})\}$. The effective radius of the structure is $R_{eff}\sim 1.15m_s^{-1}$, and the total energy in a shell of $R=4$ is $E_{R\leq4}\sim 6.9$. Note that the majority of the energy density is nearly spherically-symmetric, although there is some time-dependent wobbling which is more pronounced at lower energy isosurfaces. } 
\end{figure}

\begin{figure} 
\includegraphics[scale=.6,angle=0]{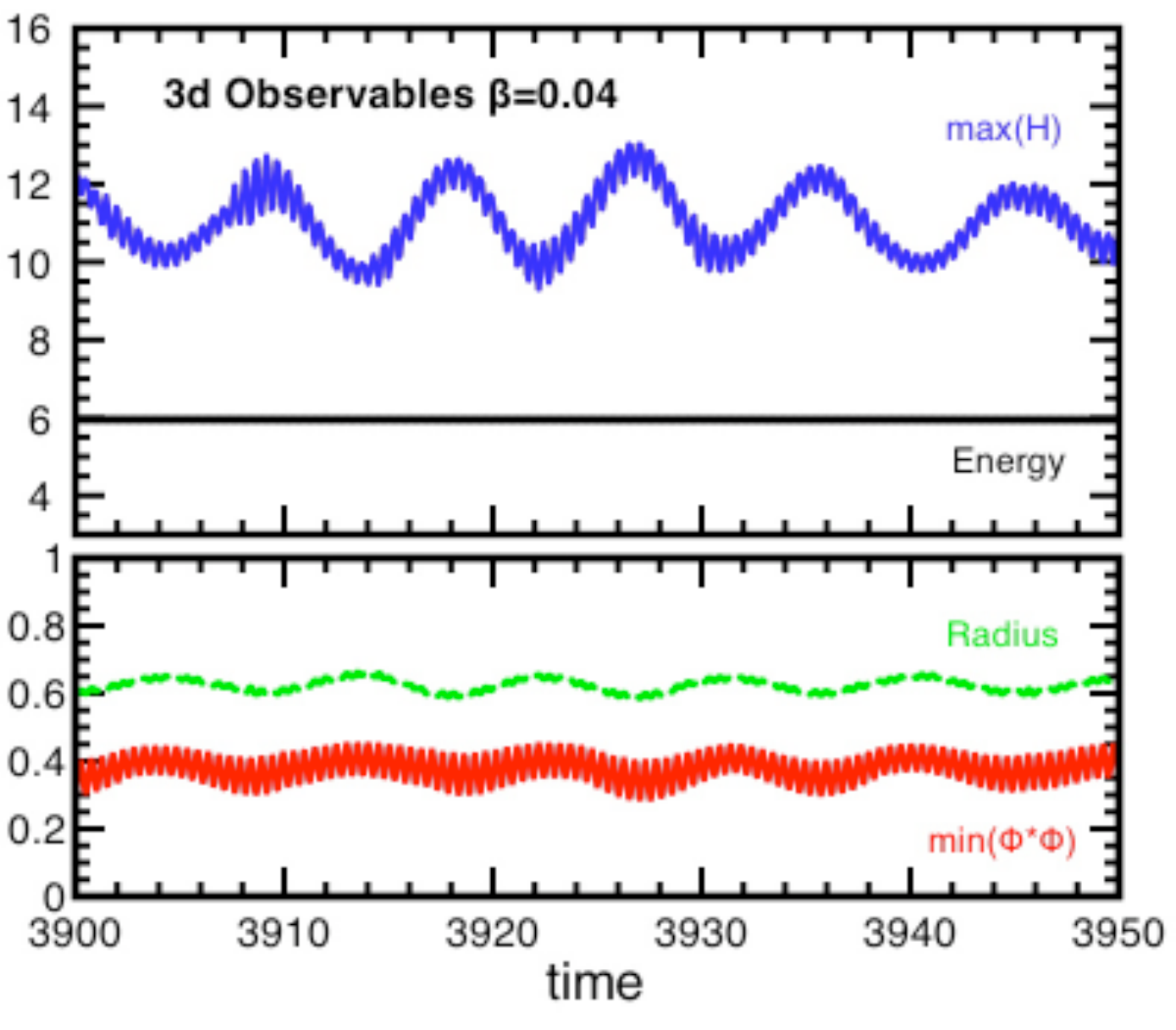}
 \caption{\label{fig:3d_oscil} (Color online) A few observables characterizing the $3d~ U(1)$ oscillon. The blue line is the maximum energy density of the EFC while the black line is the total energy within a radius of $R=4$. The lower plot shows both the expected radius of the EFC $R_{eff}^2\equiv \frac{\int r^2 \H(r) dr}{{\int \H(r) dr}}$ (green) and the minimum value of $\P^*\P$ within the EFC (red). Note that, contrary to the $2d$ case, the scalar field amplitude probes well beyond the inflection point of the double-well potential.}
\end{figure}

To better investigate the properties of these oscillons, even if only preliminarily, we use a simple initial condition inspired by our $2d$ work on vav annihilation \cite{Gleiser2007}. In that work, we placed a vortex and an anti-vortex nearly at rest and at a short distance away from each other. The vav pair interacts, attempts to annihilate and, depending on the parameters, forms a very long-lived gauged oscillon. Once this happens we put a friction wall a certain radius away from the oscillon to damp any outside energy in radiative modes that may interfere with its motion across the lattice. As we have seen in the previous sections, this method allows the determination of the EFC's energy and structure.

In $3d$, our initial setup is similar since we can extend the field in the $\hat z$ direction without affecting the dynamics. The oscillon that forms from a flux anti-flux tube annihilation is just like the $2d$ one except for the $\hat z$ symmetry. To break the $\hat z$ symmetry we then introduce a spherical friction wall which effectively pinches off and dissipates the energy density at two polar regions away from the soon to be center of the $3d$ oscillon. Essentially, we are repeating the procedure of section \ref{section:findingosc} but implementing the spherical friction wall early so as to not only absorb radiative modes, but to break the $\hat z$ symmetry and catalyze decay into the spherical oscillon.

Since the stability of a time-dependent EFC is inherently linked to the exchange of energy between conjugate momentum ($\Pi$) and potential energy ($V[\P]$), adding friction at the edges catalyses its decay. An example of an oscillonic flux tube decay catalyzed by friction is shown in figure \ref{fig:backbone} where a flux tube of length $12.5m^{-1}_s$ is pinched by a centered wall of friction of diameter $d=8m^{-1}_s$. We display snapshots of both the scalar-field condensate amplitude (red isosurface at $\P^{\dagger}\P=0.5$) and the electric field contributions to the energy density (light-blue isosurface at $\frac{1}{2}\tilde{E} \cdot \tilde{E}=0.03$). Notice how the scalar field condensate oscillates between tube-like and bubble-like configurations, while the low-energy electric field alternates between complex oscillatory patterns. 
From this figure, it is clear that the path to oscillon formation is far from trivial, even if the final configuration will display, as we will show, near-spherical symmetry. 

It should also be noted in passing that the number of (nearly) independent EFC's that form is sensitive to the length of the initial flux tube and to where we pinch it. If, for example, we were to take a longer line, then, as shown in figure \ref{fig:birth}, nearly-spherical oscillons would form at the edges, connected by a flux tube between them. Perturbations during the formation process propagate down the structure creating an interesting pattern of oscillating magnetic and electric fields. Looking at the lower isosurface plot in figure \ref{fig:birth} we see that the magnetic and electric fields create a chain-like structure that is oscillating in both time and space. This chain then clumps into a certain number of oscillons per unit length. Although quite interesting, we do not pursue the study of these hybrid objects any further.

Back to the oscillon structure with $\b=0.01$ of figure \ref{fig:backbone}, once the flux-antiflux line oscillon relaxes into the approximately spherical oscillon and we have also absorbed much of the external radiative and thermal modes, we can look at the fine structure of the low energy fields to investigate the mechanism by which energy is being slowly radiated at large times, as showed in figure \ref{fig:radiation}. From figure \ref{fig:radiation}, we note that at large times the energy within a shell surrounding the oscillon can be written as
\be 
E(t) \simeq c_0 + c_1 t^{-c_2} ~,
\ee 
where we chose a radius $R=4$ around the maximum energy density. Extrapolating to $t\rightarrow \infty$, we can see that the radiation will approach zero and there will be a finite energy left in the EFC, namely $E(\infty) \simeq c_0$. Inspecting the nonspherical fine structure of the energy density, we find that the surviving large time structure is very nearly spherically symmetric and that a small fraction of its energy is bound up in a combination of oscillatory modes in the electric and magnetic fields which resemble transitions between excited atomic states.

From the continuing snapshots of figure \ref{electric_sphere_harmonics}, we see that the energy density in the electric fields resembles a superposition of alternating mixed spherical harmonics. As time increases, the amplitude of the higher harmonics decreases and the energy in the electric field becomes more spherically symmetric (s-orbitals dominate). If we look at the magnetic field contribution, as in figure \ref{magnetic_sphere_harmonics}, we observe that the low energy component assumes a toroidal shape (see $t=202m^{-1}_s$ plot from figure \ref{magnetic_sphere_harmonics}). Because the oscillon is intrinsically time dependent, the actual asymptotic profile of the magnetic energy $B(x)\cdot B(x)$ is unclear, although we do see signs of a toroid with a continual precession as well as an oscillation in amplitude. We stress, however, that these remarks are very preliminary and that a more detailed analysis is still lacking.

It is also important to stress that these fine structure harmonics only account for less than a percent of the actual energy of the EFC. This can be seen by considering that the isosurfaces in electric and magnetic field energies are plotted at a radius of $\sim 2.5$, while the effective radius of the structure, as calculated by the ratio 
\be R_{eff}^2 \equiv \frac{\int r^2 \H(r) dr} {\int \H(r) dr} 
\ee 
is much smaller ($R_{eff}\sim 0.6$ for $\b = 0.04$ for example). In figure \ref{energy_various_isosurfaces}, we plot various total energy-density isosurfaces at late times: violet at $\H=5$; magenta at $\H=0.25$; blue at $\H=0.01$. It is clear that most of the energy is spherically-symmetric and concentrated in a very small radius, although there are departures from sphericity at large radii.
We also present the time-dependent observables for the $3d$ configuration in figure \ref{fig:3d_oscil}. This figure is to be compared with the $2d$ version for the same parameters in figure \ref{fig:2d_oscil}. Note how in $3d$ the scalar field amplitude probes well within the $V'' <0$ part of the double-well potential.

\section{An Initial $SU(2)$ $3d$ Low $\b$ Search}

In the context of non-Abelian models, the most distinctive searches for oscillons were performed in references \cite{Farhi2005} and \cite{Graham2007}. Using {\it ansatze} that approximate a spherically-symmetric oscillon, the authors found a stable oscillon in the bosonic sector of the full $SU(2)XU(1)$ electroweak model in $3d$ for the particular mass ratio $m_{\rm Higgs} = 2 m_{W}$. Given that this value of the Higgs mass is well within the reach for the upcoming LHC, the result has considerable relevance. Interestingly, any deviation from this mass ratio causes the structure to destabilize. We note that this mass ratio corresponds to type-$2$ behavior in superconductor phenomenology. 

In this section, we present an initial search in the type-$1$ region where the vector particle is much more massive than the scalar. Of course, this is not where we expect to see a scalar field in the Standard Model. However, given that non-Abelian models are an integral part of any extension to the Standard Model, or of higher level unification, it is important to map the possible nonperturbative structures that might emerge from large fluctuations about the vacuum. Only a few years back the Higgs itself could still have been much lighter than the $W$-boson. Based on our $U(1)$ work, the type-1 region is where we expect a coherent object to exist. Obviously, the non-Abelian theory is significantly different as there are no topological or even quasi-topological structures which carry as much energy as the flux tubes in the $3d$ $U(1)$ theory. This presents a challenge to our method as there are no flux tubes to lock energy in the initial part of the simulation.

\begin{figure} \centering
\includegraphics[scale=.6,angle=0]{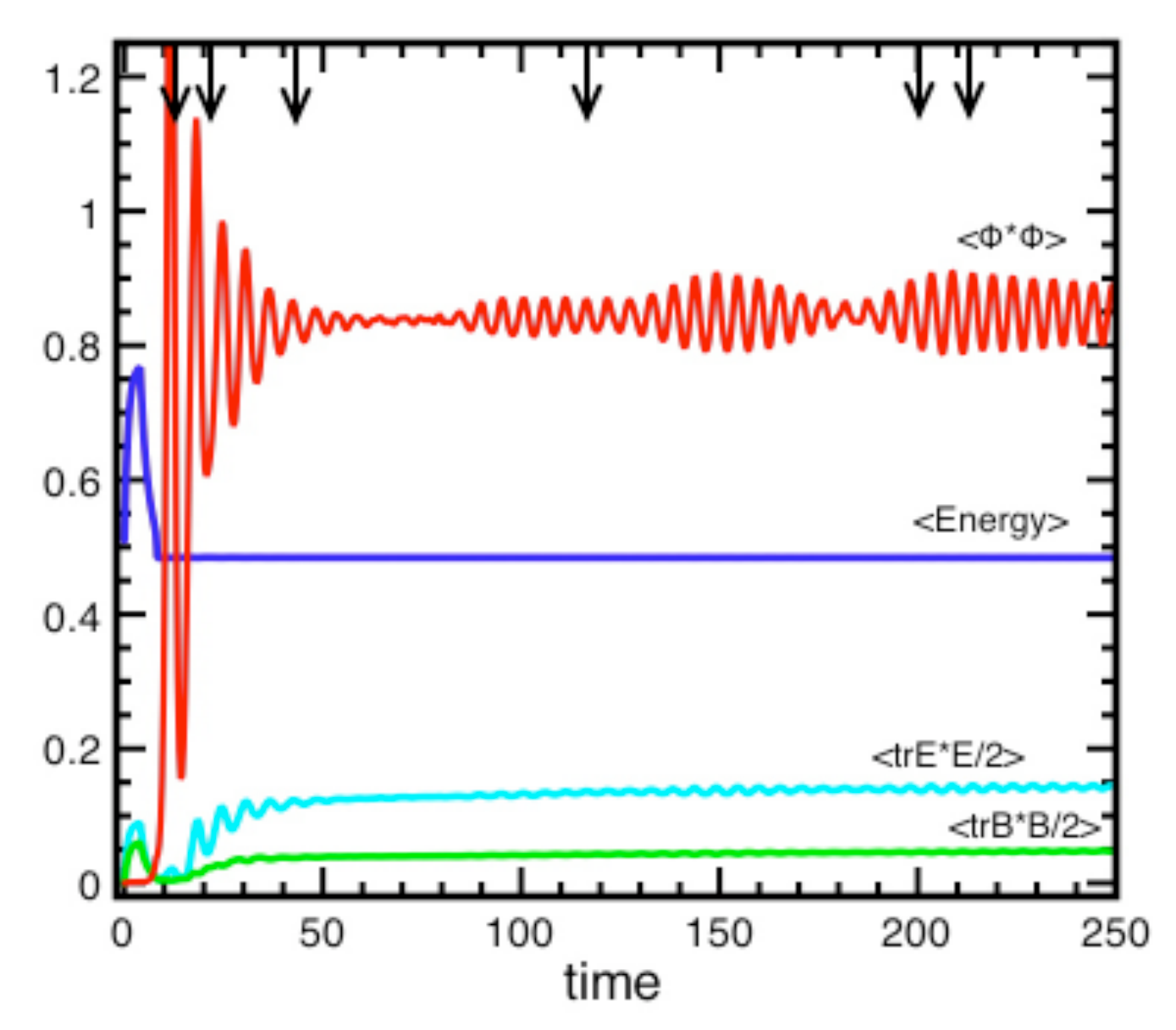}
 \caption{\label{fig:su2_av} (Color online)  Global observables for a $SU(2)$ quench at $g=4.0$ on a lattice of volume $28.8^3 m^{-3}_s$. The black arrows correspond to the time-slices in figure \ref{fig:su2em}.  The system is thermalized on a Higgs potential for $t=5m_s^{-1}$ and then cooled with $\g=1$ [CHECK] for the same amount of time.  After that, the system is evolved conservatively. $\bra \P^\dag\P \ket$ (red), $\bra \H\ket$ (blue), and the electric and magnetic components of energy are plotted in cyan and green, respectively. These colors correspond to the isosurfaces of figure \ref{fig:su2em}. } 
\end{figure}

\begin{figure*} \centering
\includegraphics[scale=0.525,angle=0]{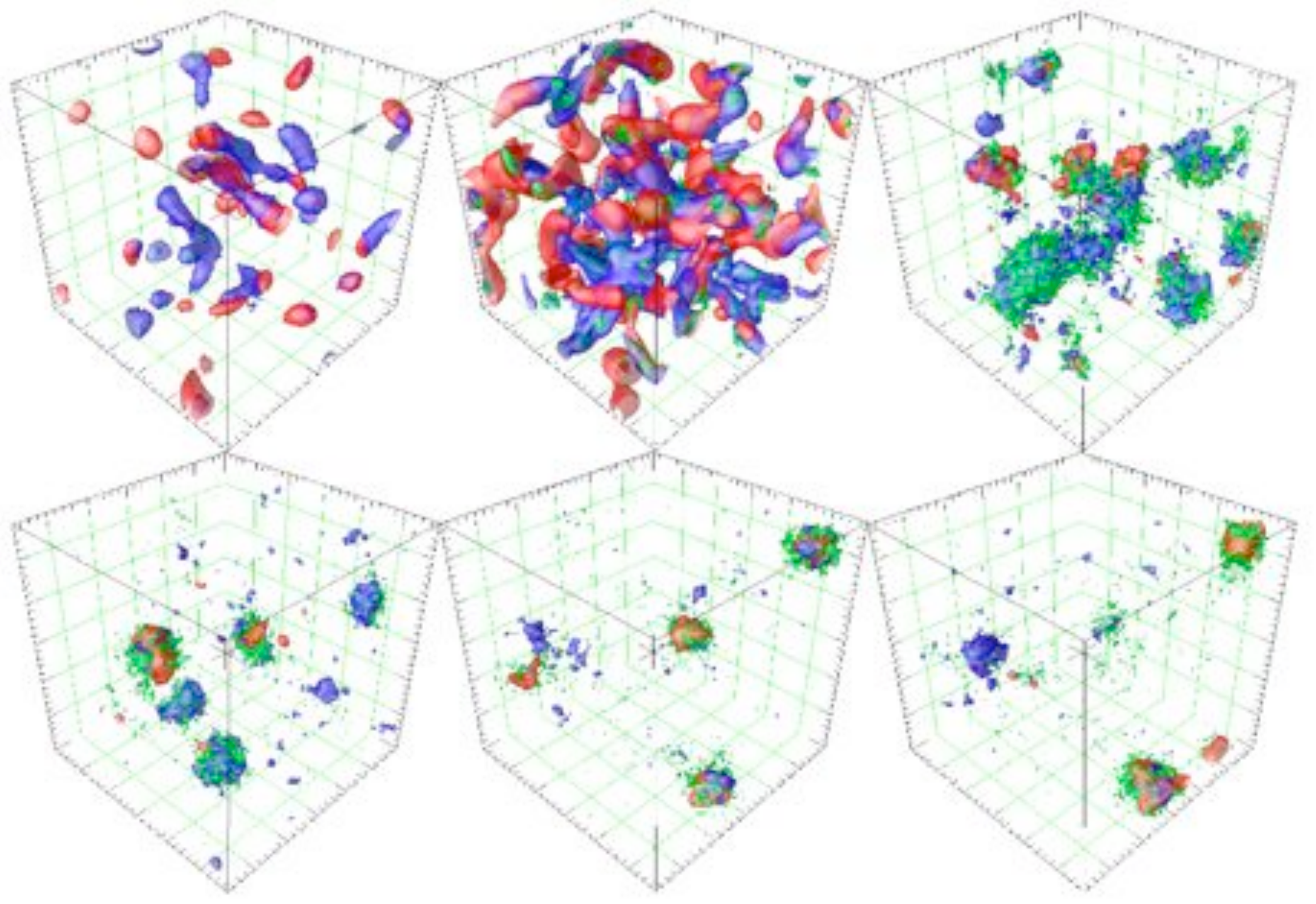}
 \caption{\label{fig:su2em} (Color online) Local observables for a $SU(2)$ quench at $g=4.0$ on a lattice of volume $28.8^3 m^{-3}_s$.  Time slice isosurfaces corresponding to the times labeled in figure \ref{fig:su2_av}.  Isosurfaces are twice their global averaged values, except for $\P^\dag \P=0.5$.  These particular configurations are large amplitude and decay quickly due to interactions with the propagating modes surrounding them. The full simulation can be viewed {\tt \href{http://media.dartmouth.edu/~mgleiser/SU2.mov}{here.} }   } 
\end{figure*}

Instead of using excessive dissipation to clean up the system during the initial moments of the simulation, we will go another route. To generate spatiotemporal complexity, we start the scalar fields at the unstable symmetric point of the potential of eq.  \ref{eq:Lag}, adding only small-amplitude thermal perturbations to excite the various modes. As the fields evolve, they probe the unstable (spinodal) portion of the potential. The amplified instabilities -- which can be viewed as isosurfaces of the energy density -- create localized quasi-bubbles connected by quasi-strings. In Figure \ref{fig:su2_av} we show a few observables: the red curve corresponds to the volume-averaged amplitude of the $SU(2)$ Higgs field which, after the initial large-amplitude fluctuations, oscillates about $\langle \P^{\dag}\P\rangle \sim 0.85$. The blue curve corresponds to the total energy and the cyan and green curves at the bottom correspond to the ``electric'' and ``magnetic'' portions of the energy, respectively. It is clear that we have achieved good energy conservation in our simulation. The vertical black arrows in the figure correspond to the time snapshots shown in figure \ref{fig:su2em}. For clarity, the colors for the isosurfaces follow the conventions of figure \ref{fig:su2_av}.
As can be seen in the figures, the complex network settles into a few large-amplitude (${\rm max} [\P^*\P] \sim 0$) high-energy localized objects, which then decay at about $t\sim 400m^{-1}_s$. Visualizations of the simulations can be seen {\tt \href{http://media.dartmouth.edu/~mgleiser/SU2.mov}{here.} }

The fact that these structures live only to $t\sim 400m^{-1}$ says very little about their  stability in near-vacuum. Since we have dumped so much energy into the system, there is a plethora of large-amplitude propagating modes which can and do disrupt oscillon stability. We have seen this in the context of $U(1)$ models and in models with only real scalar fields: oscillons that have a very large lifetime {\it in vacuo} ($t\sim 10^4 m^{-1}_s$ ), if put into a high temperature background can have their lifetimes drastically reduced to $t < 10^3 m^{-1}_s$. This is not surprising in situations where oscillons, which are coherent field configurations, must do work against an incoherent background \cite{Gleiser1996}. So, it would be necessary ``to clean up'' the lattice which have these emergent configurations in order to ascertain if there is a really long lived $\sim$type-$1$ $SU(2)$ oscillon. We note that in either laboratory or cosmological applications, symmetry-breaking is accompanied by a quench or cooling which will help oscillons -- if present -- to survive. We plan to  continue this study and to investigate if there are any configurations which are long-lived outside of the special $2:1$ mass ratio observed by ref. \cite{Farhi2005}.

\section{Summary and Outlook}

We have presented a numerical study of symmetry-breaking in the $3d$ Abelian-Higgs model. In particular, we searched for nontopological, time-dependent, long-lived configurations that may emerge dynamically as the system relaxes to the lower-energy asymmetric vacuum. Our results indicate that such oscillon-like solutions to the equations of motion can be easily found for a wide range of parameters. For the $U(1)$ model, the control parameter is $\b=(m_s/m_v)^2$. We found that oscillons exist within the type-1 regime, for $\b < 0.09$. It remains to be seen if this range can be extended for larger $\b$. One shortcoming of our approach is that it dumps too much energy in the initial state, as we quench from the symmetric to the broken symmetric vacuum. It should be possible to extract the approximate field behaviors that characterize oscillons from our numerical solutions and use them as {\it ansatze} in cleaner searches as, say, was done in refs. \cite{Copeland1995} or \cite{Farhi2005}. This way, we could probe into the type-2 regime in search of oscillons. Our preliminary study of their rich resonances and of possible oscillons in type-1 $SU(2)$ models suggest that there is much more to be explored. Although analytical results are very challenging, it is possible that some may be achievable extending the procedure of ref. \cite{Gleiser2008} to models with gauge fields.

The fact that oscillon-like configurations emerge spontaneously during symmetry breaking should not be overlooked. There is very poor understanding of the dynamical aspects of symmetry breaking in gauge theories as they involve complex nonlinear and possibly nonequilibrium physics. The emergence of oscillon-like EFCs in the Abelian-Higgs model indicates that the thermodynamics of these systems is far from trivial. In cosmology or in colliders, the presence of these configurations will delay equipartition and thus the final approach to equilibrium. They may, for example, affect the calculation of the reheating temperature in inflationary models \cite{Bassett2006}. They may also affect the structure of the vacuum and thus the computation of transition amplitudes between vacuum states \cite{Weinberg1996}. We are now entering an era where even desktop computers have enough power to perform such simulations. It is clear that much new physics remains to be explored.

\acknowledgements
MG and JT were partially supported by a National Science Foundation grant PHY-0757124. We also would like to thank  the NCSA Teragrid cluster for access under grant number PHY-070021.


\appendix
\section{Lattice Implementation}

In order to construct gauge invariant objects in a Hamiltonian lattice formulation we will be using the Wilson loop formalism \cite{Wilson1974}.
For the lattice Laplacian, we use the second-order version of the covariant derivative \cite{Smit2001}
\be \label{dlatt} \D^2_\mu \P(x) \sim \frac{1}{\dx^2} \lb U^\mu_{\hat x} \P_{\hat{x}+ \hat{\mu}} + U^{\dagger\mu}_{\hat{x}-\hat{\mu}}\P_{\hat{x}- \hat{\mu}} -2\P_{\hat x} \rb +\O(\dx^2),
\ee
where the hatted variables ${\hat x}$ denote spacetime lattice positions and ${\hat \mu}$ denote a lattice displacement by one lattice spacing $\delta x$. 
The gauge field strengths are constructed from the unitary link variables $U(x,\mu)$ satisfying the lattice gauge transformations
\be 
\label{linktrans} U(x,\mu)\equiv U^\mu_{\hat x} \rightarrow \Lambda_{\hat x} U^\mu_{\hat x} \Lambda_{\hat {x} + \hat{\mu}}^\dagger.
\ee
We can use the link variables to construct the unitary tensor
\be 
\label{unit_tensor} U^{ij}_{\Box}(x) \equiv U^i U^j_{+\hat{i}} {U^i_{+\hat{j}}}^\dagger {U^j}^\dagger, 
\ee  
which is the fundamental lattice gauge invariant object from which to derive the field strengths. Here, ${i,j}$ represent spatial indices. The magnetic contributions to the Hamiltonian can be found from the combination 
\be 
\frac{1}{2} \tilde{B} \cdot \tilde{B} = \frac{1}{4}F^{ij}\cdot F_{ij} \simeq \frac{1}{g^2 \dx^4} \lb 1-\re \frac{\Tr(U^{ij}_{\Box})}{\r} \rb, 
\ee 
where $\r=2$ for $SU(2)$ and $\r=1$ for $U(1)$. The field strength is  
\be 
F^{ij} \simeq \frac{1}{g \dx^2} \im U^{ij}_{\Box}. 
\ee 

The scalar field will be a nonunitary complex matrix which transforms as \be \P(x) \equiv \P_{\hat x} \raw \Lambda_{\hat x} \P_{\hat x}. \ee 

The covariant derivative of the force tensor needed for the equations of motion can be found from the variation, 
\be \frac{\d}{\d A_\mu(y)} \sum_x U^{\mn}_{\Box}(x). 
\ee 
The same result can be obtained from constructing the covariant derivative of the field strength in the adjoint representation by taking a gauge invariant backward derivative on the field strength tensor, 
\be 
\D_\mu F^{\mn} \simeq  \frac{1}{g \dx^3} \im \lb U^{\mn}_{\Box}(x) -  {U^\mu_{\xmm}} U^{\mn}_{\Box}(x-\hat{\nu}) {U^\mdag_{\xmm}}  \rb, 
\ee 
where for an Abelian theory the extra ${U^\mu_{\xmm}}$'s drop out to unity. 

The momentum components are just simple complex matrices at each point in space for the scalar, and a vector of complex nonunitary matrices for the electric fields. Since we are in a gauge in which the temporal vector potential is set to zero, derivatives with respect to time are simply, $E_i=\pd_t A_i$. We can then write the gauge equations of motion on the lattice as, 
\bea 
\label{eomE} \pd_t E_\hx^j &=& \D_{i} F^{ij}_\hx + J_\hx^j \\
\pd_t A_\hx^j &=&E_\hx^j, \label{eomU_}
\eea
where the current is 
\be 
J^\mu =  \frac{2 g}{\dx} \im \left(  \P^\dagger_{\hat x} U^\mu_{\hat x}  \P_{\hat{x} +\hat{\mu}}  \right). 
\ee
 
This equation is now in a perfect form for our higher-order symplectic integration scheme, and works also with the link formalism, provided that we change equation \ref{eomU_} to the form, \be U^j(x,t_{+})=U^j(x,t) e^{-ig\dx\dt E^j(x,t)}. \ee 
That is, provided our numerical exponential is accurate (and the electric field stays within its group during simulations with stochastic forcing, see below) then the links stay unitary at all times. The gauge condition $A_0=0$ giving the time component of equation \ref{fmunu_} becomes a nondynamical constraint equation, the Gaussian constraint,
\be 
\label{gauss__} \frac{1}{\dx} \sum_{i} ( E^i_\hx - E^i_{\hx -\hat{i}} ) = 2 q \im (\Pi_x^\dagger \P).  
\ee 
Provided that  this is satisfied initially, then it is maintained to numerical accuracy at all times for the $U(1)$ theory. There are some issues in exact convergence for the non-Abelian theories.

The scalar equations are also split into conjugate momentum fields to give \bea  \pd_t\Pi_\hx  &=& \D_{latt}^2 \P_\hx- \P_\hx ( \P_\hx^\dagger \P_\hx -1 ) \\  \pd_t \P_\hx &=& \Pi_\hx  \eea where $\D_{latt}^2\P$ is defined in eq. \ref{dlatt}.
The Hamiltonian density $\H_\hx$ that corresponds to these equations is then given by,  \begin{widetext}  \be \H_\hx = \Pi_\hx^\dagger \cdot \Pi_\hx + \D_i \P_\hx^\dagger \cdot \D_i \P_\hx  + \frac{1}{2} E_i \cdot E^i +\frac{1}{g^2\dx^4} \sum_{i\neq j} \lb1- Re( \frac{\Tr(U^{ij}_\Box)}{\r}) \rb + \frac{1}{2} \left( \P^\dagger_\hx \cdot \P_\hx -1 \right)^2 .\ee  \end{widetext}

The \La implementation of these equations is explained in the next two appendices.
\section{Gauge \La Implementation $1$: Forcing Momenta}  \label{app_mom_per}

In this appendix we summarize the two methods of implementing stochastic noise and friction terms in systems with local gauge symmetries and thus Gaussian constraints. The methods differ in that whereas one couples noise to the fields, the other couples noise to the conjugate field momenta. We also explain how the implementation which forces the fields can be made to satisfy the Gaussian constraint equation at all times. Conversely, the method of forcing the momenta destroys this constraint and thus the system must be ``stopped,'' (that is, one must set all momenta set to zero) before the true simulation (symmetry breaking) begins. 

It is useful at times to run simulations with only the dissipative terms present and no thermal noise in order to relax to a ground state or specific configuration. We also present a proof which shows that spatially homogeneous friction does not affect the Gaussian constraint and is safe to use in a simulation.

Consider a $U(1)$ gauge field coupled to a complex scalar field with some Hamiltonian $H=\sum_x \H$ which is a function of the conjugate momentum fields for the scalar and the vector, $\pi^\a \equiv \lb \Pi, \tilde{E} \rb$, and the respective fields $\p^\a \equiv \lb \P, \tilde{A} \rb$. The \La dynamics for this system is, \bea
\pd_t\pi^\a_x + \g \pi^\a_x &=&-\pd_{\p_x} \sum_x \mathcal H(\pi^\a_x,\phi^\a_x) + \z_{\pi^\a} \\ \nonumber
\pd_t\p^\a_x&=&\pd_{\pi_x} \sum_x \mathcal H(\pi^\a_x,\phi^\a_x)
\eea where  $\zeta=\sqrt{\frac{2 \g T}{\dx^d \dt}} \rm{rnd_G}()$, where $\rm{rnd_G}()$ is a Gaussian random number of zero mean and unit variance. Since $\pi^0=\Pi$ is a complex number, we must add a random kick for each degree of freedom and thus $\z_{\pi^0}=\z_{\re} +i \z_{\im}$, where both the real and imaginary components of $\z_{\pi^0}$ are identical except for their random numbers. Numerically, they are just sequential calls to the random number generator. If we had written the set of equations as a scalar field of $O(n)$ symmetry then we would just call the random number $\z$ $n$ times and couple that force linearly to each component of the $O(n)$ field. This is also the same for the vector field as each $\pi^{i}=E^{i}$ component gets its own random kick. 

This \La evolution violates locally the Gauss constraint, 
\be 
\label{apBgauss}  
\frac{1}{\d x} \sum_i (E^i_{x-i}-E^i_x)=i(\pi^0_x(\p^0_x)^\dagger -(\pi^0_x)^\dag \p^0_x).    \ee
Since this equation is preserved by the equations of motion without the forcing noise, to satisfy it we can just set all the momenta to zero at the beginning of a simulation. Although this procedure also destroys the thermal state, it will approximately be restored after the energy is quickly transfered from the spatial derivatives to the temporal ones as the system tries to reach equipartition. We stress that we are not interested in enforcing a strict thermal state anyway, only in using the Langevin approach to excite a large spectrum of field modes.

Note that the $\g$ frictional coefficient is not a problem as it scales smoothly in $k$-space as $e^{-\g t}$ and so does not change any spatially local term. We can show that this has no effect by considering the equations \bea \pd_t \pi +\g \pi &=& \D^2\p - m^2_{eff}\p \\ 
\pd_t E^i +\g E^i &=& \pd_i F^{ij} + J^i \\ 
\pd_i E^i &=& J^0.  \label{derv01}
\eea Taking the divergence of the second equation and the time derivative on the last and then subtracting the two gives: 
\bea 
0+\g \pd_i E^i &=& 0 + \pd_i J^i-\pd_t J^0.  \label{derv02}
\eea 
In order to get the appropriate correction that should cancel the $\g$ term we expand \bea
 \pd_t J^0 &\raw& 2g\im \pd_t \pi^\dagger \p \\ &\raw& 2g \im ((-\g \pi^\dagger +\D^2 \p^\dagger -m^2_{eff} \p^\dagger)\p) \\ &\raw&
\pd_i J^i -2g\g \im \pi^\dagger \p.
 \eea 
So equation \ref{derv02} can be expressed as 
\bea 
\g \pd_i E^i &=& \g J^0,
\eea 
which means that the continuity equation is of the same form as the initial $\g=0$ constraint equation \ref{derv01}.

\section{Gauge \La Implementation $2$: Forcing Fields} \label{app_den_per}

Recently we became aware of the work by Krasnitz on how to maintain constraints on gauge fields during \La evolution \cite{Krasnitz1995}. After some experimentation with the algorithms in reference \cite{Krasnitz1995}, we were able to find a simpler and less computationally demanding method of maintaining the Gaussian constraint on the $U(1)$ gauged scalar system. While not as general as the methods of \cite{Krasnitz1995}, for at least a subset of the Abelian systems our method is simpler and numerically faster Langevin-type thermalization method. In Krasnitz's method, the force function must be called twice per time step and the simulation time almost doubles. 

For clarity, we repeat the basic equations for an Abelian system with friction, \bea \pd_t A_i &=& E_i \\  \pd_t \P &=& \Pi \\ \pd_t E_i &=& \pd_j F^{ji} -\g E_i + 2g\im \P \D_i \P^* \\ \pd_t \Pi &=& \D^2 \P - \g \Pi - V'[\P]   \eea and the Gaussian constrain equation 
\be 
\pd_i E^i = 2g \im  \Pi^* \P. \label{B_gauss}, 
\ee 
where it is implied that we use the lattice version of the operators. 

The method works as follows: instead of adding a stochastic force which satisfies a fluctuation-dissipation relation to the momentum evolution equations, we add a random  perturbation to the field equations. Thus, the amplitude perturbations are written as
\bea 
\pd_t A_i &=& E_i + \z_E \\  \pd_t \P &=& \Pi +\z_\Pi.  
\eea 
It is obvious that a change in amplitude of $A_i$ cannot affect the electric field and so the divergence in equation \ref{B_gauss} is not modified. Let us look at the charge density at two separate times to see what conditions on $\z_\Pi$ will maintain the right side of the constraint equation \ref{B_gauss}. Writing $\P(t+\dt)\equiv\P_+ = \P + \Pi \dt +\z_\Pi \dt$, gives the charge density at half a leapfrog step,
\bea 
J^0_+(\P_+) &=& J^0( \P + \Pi \dt +\z_\Pi \dt) \\ 
2g \im \Pi^* \P_+ &=& 2g \im \Pi^* (\P + \Pi \dt +\z_\Pi \dt) \\
&=& J^0 + 2g \dt \im \Pi^* \z_\Pi,
\eea 
where we only need to advance $\P\raw \P_+$ because of the leapfrog scheme of integration. If $J_0$ is locally conserved at this half step then although it is modified in the next half step, it is not changed in any way which violates equation \ref{B_gauss}. Withis we can get the necessary form of $\z_\Pi$ which makes $\im \Pi^*\z_\Pi = 0$ and thus maintains the local Gauss constraint. If we choose a real stochastic variable $\z$ and define 
\be 
\z_{\Pi} \equiv \frac{\Pi}{\sqrt{\Pi^*\Pi}}\z, 
\ee 
we can add random fluctuations at any time during the simulation without having to set all momenta to zero to regain the Gaussian constraint as we must with the first type (momentum forcing) of implementation.

It should be noted that this particular implementation will not randomize the phase of an ungauged complex scalar, since it ends up being only a density perturbation which relies on the gauge field to randomize the phase. So to thermalize a complex scalar, use a complex $\z=\z_R+i\z_I$ instead of $\z_{\Pi}$.

 \subsection*{Criticisms and Challenges}
 
There is one obvious criticism to the thermalization technique presented in this section. If we look at the equations, we see that we are in effect only modifying the spatially-dependent degrees of freedom by making fluctuating charge and magnetic densities but not doing anything that would make the charge-anticharge and electric distributions come to equilibrium. 

For example, in reference \cite{Krasnitz1995} the forcing does change both sides of equation \ref{apBgauss} but in a way that is balanced. This seems a better way of doing things at the cost of a few more computational steps, as it does not change the equilibrium configuration that the forcing directs the system towards. Just because we are not modifying either side of equation \ref{apBgauss} does not mean that the charge densities remain trivial. Energy flows very quickly and effectively from the current to charge-anticharge and from magnetic to electric degrees of freedom inherently because of the equations of motion. So, a thermal equilibrium distribution of $\pd_i E_i$ and spatially varying charge densities does in fact form.

Another issue to be careful of is that when thermalizing from the amplitude perturbations, the largest one can make the electric contributions to the temperature is $\bra E_i E^i \ket = \bra B_i B^i \ket$, which comes from minimization of the Lagrangian instead of distributing $\frac{T}{2}$ per degree of freedom in the Hamiltonian. 

This happens anyway if we maintain the Gauss constraint by setting all the momenta to zero. None of these techniques give a perfect thermal distribution.

With both techniques we see that generally the momenta thermalize well $\bra \Pi^* \Pi \ket \propto \bra E_i E^i \ket \propto T$ as do the surface kinetic terms $\bra \pd_i\P^* \pd^i \P \ket$. But the energy mixture can have issues between the magnetic and the electric terms.  These issues are present in every form of real time classical thermalization that we have tried.

\end{document}